\documentclass[fleqn,usenatbib]{mnras}

\usepackage{newtxtext,newtxmath}

\usepackage[T1]{fontenc}

\DeclareRobustCommand{\VAN}[3]{#2}
\let\VANthebibliography\thebibliography
\def\thebibliography{\DeclareRobustCommand{\VAN}[3]{##3}\VANthebibliography}

\usepackage{booktabs}
\usepackage{tabularx}
\usepackage{bigdelim}
\usepackage{lipsum}
\usepackage{xcolor}
\usepackage{graphicx}	
\usepackage{amsmath}	
\usepackage{orcidlink}
\usepackage{float}

\newcommand{\galform}{\textsc{galform}}

\title[MW satellites: sensitivity analysis]{A galactic tug-of-war: how (not) to simultaneously fit the Milky Way satellite luminosity function {\it and} the mass-metallicity relation}

\author[S. Bose \& A. Deason]{
Sownak Bose\orcidlink{0000-0002-0974-5266}$^{1}$\thanks{E-mail: sownak.bose@durham.ac.uk (SB)},
Alis J. Deason\orcidlink{0000-0001-6146-2645}$^{1}$
\\
$^{1}$Institute for Computational Cosmology, Department of Physics, Durham University, South Road, Durham DH1 3LE, UK\\
}

\date{Accepted XXX. Received YYY; in original form ZZZ}

\pubyear{\the\year{}}

\begin{document}
\label{firstpage}
\pagerange{\pageref{firstpage}--\pageref{lastpage}}
\maketitle

\begin{abstract}
The satellite population of the Milky Way is shaped by a range of astrophysical processes including mergers, star formation, feedback, and cosmic reionisation. Determining which processes most strongly influence its properties is challenging and a key test of galaxy formation models. We train a neural network on the \galform{} semi-analytic model and apply a variance-based sensitivity analysis to characterise the influence of 11 astrophysical parameters on two key observables: the satellite luminosity function, and the mass-metallicity relation. We find that: (1) the abundance of bright satellites ($\mathrm{M}_V \lesssim -13$) is regulated  by supernova feedback; (2) the faint end of the luminosity function is shaped by the interplay between feedback and reionisation; and (3) the mass-metallicity relation is governed almost exclusively by feedback at all masses. We do not find a combination of parameters in the fiducial model that fits the observed data for both statistics simultaneously. To understand why, we employ SHapley Additive exPlanations to capture the {\it directionality} of each parameter variation. This enables us to pinpoint the origin of tensions in the model, showing that parameter adjustments that regulate the abundance of faint satellites drive stellar metallicities to be an order of magnitude too low and vice versa. This ``tug-of-war'' leads us to consider extensions to the baseline model, such as metal loading in winds, or allowing the feedback strength to evolve with redshift. Our study highlights the value of interrogating complex physical models through a sensitivity analysis framework by revealing high-order parameter interactions and non-linear responses that traditional one-at-a-time variations would miss.
\end{abstract}




\section{Introduction}
\label{sec:intro}

In our present paradigm of structure formation, the so-called $\Lambda$ Cold Dark Matter ($\Lambda$CDM) model, structures grow through hierarchical mergers of dark matter clumps, the aggregation of which results in a population of dark matter haloes with a mass spectrum that spans several orders of magnitude \citep{Frenk2012,Wang2020}. These haloes are the sites of galaxy formation, within which gas cools and condenses, before eventually forming stars \citep{White1978}. This process, however, is not perfectly efficient. The properties of galaxies are shaped profoundly by feedback mechanisms that regulate star formation. Stellar winds and supernovae explosions can heat and expel gas from the host haloes of galaxies, temporarily quenching star formation \citep{Larson1974,Dekel1986}. On yet larger scales, the cumulative ultraviolet (UV) radiation from the first generations of stars reionised the Universe, heating the intergalactic medium (IGM) and preventing gas from cooling into low mass haloes with virial temperatures below $\sim 10^4$ K \citep{Couchman1986, Efstathiou1992, Thoul1996, Gnedin2000}. The end result of these myriad processes is a galaxy population that is much smaller in number compared to the population of dark matter haloes, but which has a rich diversity of present-day properties.

The dwarf satellite population of the Milky Way is a particularly useful laboratory for testing our theoretical understanding of these processes. This is largely thanks to the discovery of increasingly faint systems through the combined effort of spectroscopic and deep imaging surveys like the Sloan Digital Sky Survey, the Dark Energy Survey, the Hyper Suprime-Cam Survey, and Pan-STARRS1 \citep[][]{Willman2005,Bechtol2015,Drlica2015}. At present, some 66 such objects have been confirmed, but the total number could be as large as several hundreds \cite[][]{Newton2018,Kim2018,Nadler2020}, with many potentially still awaiting discovery \citep{Manwadkar2022,Santos2025}. These dwarf galaxies are the remnants of the earliest structures to form and are uniquely sensitive to the physics of feedback, cosmic reionisation, and even the nature of dark matter \citep[see, e.g.,][]{Nadler2019,Newton2021,Enzi2021,Mau2022,Tan2025}. Indeed, the inclusion (or lack thereof) of appropriate processes related to baryonic physics can mean the difference between claiming tensions in the $\Lambda$CDM model, or showcasing its successes in the highly non-linear regime (see recent reviews by \citealt{Bullock2017} and \citealt{Sales2022} on this subject).

While this provides a compelling qualitative picture, a quantitative understanding remains elusive. Modern theories of galaxy formation are implemented in detailed numerical simulations or semi-analytic models, which depend on a large number of parameters that encapsulate the uncertain physics of star formation, feedback, and their interplay. For example, there is some consensus that the efficiency of supernova feedback, often represented as a `mass loading' factor, likely depends on the potential well of the galaxy. However, the exact functional dependence (i.e., normalisation and power-law index) of this relation is poorly constrained from first principles and not universally agreed upon amongst theoretical models and simulations \citep[e.g.,][]{Lagos2013,Muratov2015,Nelson2019,Mitchell2020,Pandya2021,Bassini2023} -- not to mention the challenges that come with measuring this quantity observationally \citep[e.g.][]{Chisholm2018,McQuinn2019,Romano2023,KadoFong2024}. Similarly, the manner in which cosmic reionisation impacts the galaxy population, shaping its properties into those we observe today, remains subject to theoretical uncertainty \citep[although considerable progress has been made on this front in recent years; see, e.g.,][and \citealt{Gnedin2022}, for a review]{Katz2020,Kannan2022,Zier2025,Rey2025}. The high dimensionality of the parameter space, the complex, non-linear interactions between the physical processes, and the stochastic nature of these calculations\footnote{It may well be argued that the stochasticity observed in hydrodynamical simulations is a {\it desired} outcome rather than a limitation, albeit one that comes at the expense of making certain tests hard to repeat in a controlled fashion.}, make it challenging to build an intuitive understanding of the model or to identify the root cause of its successes and failures. It is often unclear which physical processes are most responsible for shaping which observable properties of the satellite population.

Semi-analytic models of galaxy formation are highly informative theoretical tools for this kind of exploration, particularly in the regime of dwarf satellites, where the computational demands are especially acute \citep[indeed, most modern hydrodynamical simulations focus on the population of {\it isolated} dwarfs, e.g.,][]{Munshi2019,CWheeler2019,Emerick2019,Koudmani2021,Christensen2024}. Although they assume spherical symmetry, which limits their ability to fully track the evolution of gas in galaxies as in hydrodynamical simulations, semi-analytic (and semi-empirical) models are significantly cheaper from a computational standpoint. This advantage allows for the generation of a large statistical sample of galaxies and rapid exploration of the parameter space describing the physical processes implemented in the model, a feature that makes it easy to examine how model predictions are affected by turning on or off specific mechanisms -- or to model them in a different way entirely. Consequently, these models have become a key addition to the theorist's toolset for understanding the properties of dwarf galaxies in a controlled and systematic way \citep[e.g.,][]{Benson2002b,Font2011,Krumholz2012,Bose2018,Kravtsov2022,Kim2024,VWheeler2025}. Semi-analytic models aim to capture, as best as possible, the physical equations relating to the formation and evolution of galaxies that are then solved in a self-consistent way as the host dark matter haloes grow and merge (see, e.g., \citealt{Baugh2006} and \citealt{Somerville2015} for reviews on semi-analytic modelling). The free parameters of these models are typically calibrated to reproduce a small set of properties of the galaxy population, predominantly in the regime of objects much more massive than local dwarfs. As a result, any corresponding properties in the dwarf regime (such as their stellar masses, metallicities, or the fraction of haloes they occupy) can be considered {\it bona fide}, emergent predictions of these models. 

The goal of the present work is to build on this progress and perform a comprehensive analysis of the Milky Way satellite population to characterise the physical processes that shape its properties. Here, we will focus on two fundamental observables: the luminosity function, which quantifies the number of satellite galaxies as a function of stellar mass (or brightness), and the mass-metallicity relation, which encodes the chemical abundances in these galaxies. We use the Durham semi-analytic model, \galform{} \citep{Cole2000, Lacey2016}, to model the satellite population around several realisations of Milky Way-mass haloes, varying the model parameters over a wide range. To navigate the complex and high-dimensional parameter space of this model, we develop a neural network emulator that allows us to replicate the predictions of the \galform{} model for any choice of input parameters rapidly and accurately. This builds on the pioneering work of \cite{Bower2010}, who first explored the use of neural networks in the context of semi-analytic models of galaxy formation using the \galform{} model. We then perform a variance-based sensitivity analysis of our model, for which we use the method of Sobol' indices \citep{Sobol2001}. The Sobol' indices allow us to quantify the fraction of the variance in each of the model predictions that can be attributed to each input parameter individually, and through its interactions with all other parameters. This method moves beyond one-at-a-time parameter variations to build a complete, hierarchical understanding of how the model connects its physical inputs to its output predictions. Furthermore, by employing SHapley Additive exPlanations \citep[SHAP;][]{Lundberg2017}, we pinpoint not just the importance of a parameter, but also the directionality of its impact.

The organisation of this paper is as follows. In Section~\ref{sec:theory}, we describe the theoretical concepts underpinning this work, with a particular emphasis on our semi-analytic model of galaxy formation, \galform{}, and the concept of variance-based sensitivity analysis using Sobol' indices. Our main results are presented in Section~\ref{sec:main}, in which we showcase the sensitivity analysis applied to the satellite luminosity function and the mass-metallicity relation, establishing not just {\it which} model parameters are most influential, but also {\it how} they shape each of these statistics. Section~\ref{sec:discussion} contains a discussion of some of the limitations of our model. Finally, our main conclusions are summarised in Section~\ref{sec:conclusions}.

\section{Theoretical background}
\label{sec:theory}

In this section, we will introduce some of the primary tools used in this investigation, with a special emphasis on the \galform{} semi-analytic model of galaxy formation (Section~\ref{sec:galform}). We then describe the procedure for evaluating and quantifying the sensitivity of the model predictions to particular choices of input parameters (Section~\ref{sec:sobol}). Finally, we outline the design of a neural network emulator, trained on \galform{}, which is used to perform the sensitivity analysis (Section~\ref{sec:emulator}). Readers who are not interested in this theoretical background are recommended to skip to Section~\ref{sec:main}.

\subsection{The \galform{} semi-analytic model}
\label{sec:galform} 

A semi-analytic model offers a computationally inexpensive and flexible way to generate a synthetic galaxy population given a set of input parameters that define the physical processes in the model \citep[e.g.,][]{Kauffmann1993,Cole1994,Cole2000,Somerville2008,Henriques2015,Hirschmann2016}.  The model we use, \galform{}, was first presented in \cite{Cole1994} and \cite{Cole2000}. \galform{} follows the dark matter halo merger trees and populates them with galaxies by solving coupled differential equations that capture the physics of gas cooling in haloes, star formation, feedback from stars and black holes, the evolution of stellar populations, and chemical enrichment and recycling. 

The latest version of this model was described in \cite{Lacey2016}, and incorporates several features from previous versions of \galform{}, and is what we refer to as the `fiducial' model in what follows. This model includes the inclusion of a top-heavy initial mass function (IMF) in starbursts, which is necessary to reproduce the abundance of star-forming sub-millimetre galaxies \citep{Baugh2005}. Additionally, the model includes feedback from active galactic nuclei, which regulates the growth of massive galaxies \citep{Bower2006}, as well as a star formation law that depends on the abundance molecular gas within the interstellar medium \citep{Lagos2011}. \galform{} makes predictions for the broad-band luminosities of galaxies using the stellar population synthesis model of \cite{Maraston2005}, and it has been used to explore the properties of the galaxy population across cosmic time, including galaxies at the highest redshifts that have been detected by the James Webb Space Telescope \citep{Lu2025}.  \galform{} model parameters are calibrated to reproduce existing constraints on the (present-day) optical and near-infrared luminosity functions, the black hole-bulge mass relation, the HI mass function, and the fraction of early- and late-type galaxies. An important point to emphasise is that the model has {\it not} been calibrated to reproduce the properties of galaxies in the dwarf regime, which includes the satellite population of the Milky Way.

In the following subsections, we describe some of the key physical processes relevant to this work.

\subsubsection{The assembly of dark matter haloes}
\label{sec:haloes}

At the root of any model of galaxy formation is the growth and assembly of dark matter haloes. While \galform{} can be run on the outputs of halo merger trees extracted from $N$-body simulations, in this work we choose to employ an alternative mode, in which halo assembly histories are generated using a Monte Carlo approach. In particular, we adopt the method of \cite{Parkinson2008}, which is based on the extended Press-Schechter model \citep{Bond1991,Bower1991,Lacey1993}. We use an updated version of the \cite{Parkinson2008} model, which has been adjusted slightly to better match the conditional mass function of progenitors in the regime of dwarf haloes ($M_\mathrm{halo}\lesssim10^{10}\,\mathrm{M}_\odot$). The only other input to this method is the (linear) matter power spectrum, which defines the initial matter fluctuations from which haloes collapse. Although the properties of the Milky Way satellite population may be sensitive to the nature of the dark matter particle, we assume a standard $\Lambda$CDM scenario throughout. Furthermore, we fix the cosmological parameters at the WMAP-7 values derived by \citet[][see Table~\ref{tab:params}]{Komatsu2011}. This choice has no bearing on our main conclusions. 

\subsubsection{Gas cooling within haloes}
\label{sec:gas}

In \galform{}, baryonic matter within dark matter haloes is distributed across five components: hot halo gas, an ejected gas reservoir, cold disk gas, stars, and central black holes. The model for gas accretion onto central galaxies assumes that gas entering a halo is first shock-heated to the virial temperature and settles into a cored isothermal density profile, $\rho_{\mathrm{hot}}(r)$. The rate at which this gas can accrete onto the central galaxy is given by:
\begin{equation} \label{eq:accretion}
    \dot{M}_\mathrm{acc} = 4 \pi r_\mathrm{acc}^2\, \rho_{\mathrm{hot}}(r_{\mathrm{acc}}) \frac{\mathrm{d}r_{\mathrm{acc}}}{\mathrm{d}t},
\end{equation}
where $r_{\mathrm{acc}}$ is the radius within which halo gas has had time to cool (based on the cooling time) and time to fall to the centre of the halo (based on the free-fall time). This leads to two distinct accretion regimes: a quasi-static ``hot'' accretion mode when cooling is inefficient, and a rapid ``cold'' accretion mode when gas cools quickly and falls freely towards the galaxy.

A critical distinction is made for the treatment of satellite galaxies. When a galaxy becomes a satellite upon merging with a more massive halo, we assume that its associated hot gas halo is removed instantaneously via ram pressure stripping and added to the gas reservoir of the host halo. Consequently,  satellite galaxies are completely cut off from any subsequent gas accretion. 

\subsubsection{Star formation in disks,  starbursts, and disk instabilities}
\label{sec:starform}

The star formation rate in galactic disks is calculated using the empirical scaling relation of \cite{Blitz2006}, which is derived from observations of nearby star-forming disk galaxies. In this formulation, the cold gas in the disk is divided into atomic and molecular phases with the local ratio of surface densities $(\Sigma_{\mathrm{atom}}$ and $\Sigma_{\mathrm{mol}}$) at each radius in the disk depending on the gas pressure, $P$, in the midplane as:
\begin{equation} \label{eq:Rmol}
    R_\mathrm{mol} = \frac{\Sigma_{\mathrm{mol}}}{\Sigma_{\mathrm{atom}}} = \left( \frac{P}{P_0} \right)^{\alpha_P},
\end{equation}
where we use $\alpha_P=0.8$ and $P_0/k_B=17000$ cm$^{-3}$K, based on observations of nearby galaxies \citep{Leroy2008}. The pressure, $P$, is determined from the surface densities of gas and stars, as described in \cite{Lagos2011}. The star formation rate is assumed to be proportional to the mass in the molecular component only, and is given by:
\begin{equation} \label{eq:sf_disk}
    \psi_\mathrm{disk} = \nu_\mathrm{SF} M_\mathrm{mol,disk} = \nu_\mathrm{SF} f_\mathrm{mol} M_\mathrm{cold,disk},
\end{equation}
where $f_\mathrm{mol} = R_\mathrm{mol}/\left(1+R_\mathrm{mol}\right)$, and $\nu_\mathrm{SF}$ is an adjustable parameter corresponding to the depletion timescale of this molecular gas. Our fiducial choice is $\nu_\mathrm{SF}=0.74$ Gyr$^{-1}$, which is slightly higher than the empirically-determined best-fit value from \cite{Bigiel2011}. 

Starbursts may occur during galaxy mergers or disk instabilities. The outcome of a galaxy merger depends on the ratio of the baryonic mass in the satellite (cold gas and stars), $M_\mathrm{b,sat}$, to that of the central galaxy, $M_\mathrm{b,cen}$. We identify the following scenarios:
\begin{enumerate}
    \item `Major' mergers, where $M_\mathrm{b,sat}/M_\mathrm{b,cen} > f_\mathrm{ellip}$, in which case any stellar disks are destroyed and are transformed into a spheroid component; the associated cold gas collapses entirely into the newly formed spheroid.
    \item `Minor' mergers, $M_\mathrm{b,sat}/M_\mathrm{b,cen} < f_\mathrm{ellip}$, in which the stars from the satellite are added to the spheroid of the central galaxy, but the cold gas is added to the disk of the central galaxy (with no change made to its angular momentum).
    \item Mergers where $M_\mathrm{b,sat}/M_\mathrm{b,cen} > f_\mathrm{burst}$ (which includes all major mergers) trigger starbursts, in which all of the cold gas from the merging galaxies is transferred to the spheroid component and is then either consumed by star formation or ejected by supernova feedback.
\end{enumerate}
In \galform{}, $f_\mathrm{burst} \leq f_\mathrm{ellip} \leq 1$, and both are adjustable parameters. Their default values are set to 0.05 and 0.3, respectively.

Galaxies may also trigger starbursts when they undergo morphological transformations through disk instabilities. Following the model of \cite{Efstathiou1982}, we assume that disks become dynamically unstable to bar formation if:
\begin{equation} \label{eq:f_disk}
    F_\mathrm{disk} = \frac{V_c\left(r_\mathrm{disk}\right)}{\left(1.68GM_\mathrm{disk}/r_\mathrm{disk}\right)^{1/2}} < F_\mathrm{stab},
\end{equation}
where $M_\mathrm{disk}$ is the total disk mass including both stars and gas, $r_\mathrm{disk}$ is the disk half-mass radius, and $F_\mathrm{stab}$ is an adjustable parameter set to 0.91 in the fiducial model. Defined in this way, a completely self-gravitating stellar disk would have $F_\mathrm{disk}=0.61$. If at any timestep in the \galform{} run a galaxy disk satisfies the condition $F_\mathrm{disk}<F_\mathrm{stab}$, we assume that the disk has formed a bar, which then thickens and evolves into a spheroid due to vertical buckling instabilities \citep[e.g.,][]{Combes1990,Debattista2006}. We then assume that bar formation triggers a starburst that consumes any cold gas present.

During starbursts, we assume $f_\mathrm{mol}\approx1$ and that the star formation rate timescale is related to the dynamical timescale in the host spheroid \citep[see, e.g.,][]{Kennicutt1998}:
\begin{equation} \label{eq:sf_burst}
    \psi_\mathrm{burst}=\nu_\mathrm{SF,burst}\,M_\mathrm{cold,burst} = \frac{M_\mathrm{cold,burst}}{\tau_{\star,\mathrm{burst}}},
\end{equation}
where:
\begin{equation} \label{eq:tau_burst}
    \tau_{\star\mathrm{burst}} = \mathrm{max}\left[ f_\mathrm{dyn} \tau_\mathrm{dyn,bulge}, \tau_{\star\mathrm{burst,min}} \right].
\end{equation}
The bulge dynamical time is expressed in terms of the half-mass radius and circular velocity as $\tau_{\mathrm{dyn,bulge}} = r_\mathrm{bulge}/V_c(r_\mathrm{bulge})$. The parameters $f_\mathrm{dyn}$ and $\tau_{\star\mathrm{burst,min}}$ are, in principle, adjustable, but here we keep their values fixed, respectively, at $20$ and $0.1$ Gyr.

\subsubsection{Supernova feedback}
\label{sec:SNfb}

The impact of feedback from supernovae is encapsulated in the form of a standard `mass loading' factor, $\beta$, such that the rate of gas ejection from galaxies is proportional to their instantaneous star formation rate, $\psi$:
\begin{equation} \label{eq:loading}
    \dot{M}_\mathrm{eject} = \beta\left(V_c\right)\psi = \left( \frac{V_c}{V_\mathrm{SN}^{\mathrm{disk,burst}}} \right)^{-\gamma_\mathrm{SN}}\psi,
\end{equation}
where $V_c$ is the circular velocity defined at the half-light radius of the disk for disk star formation, and of the spheroid for starbursts. $V_\mathrm{SN}^{\mathrm{burst,disk}}$ are adjustable, normalisation parameters in each of the two components, and $\gamma_\mathrm{SN}$ is power-law index of this scaling, and is also a free parameter of the model. Their fiducial values are set to $V_\mathrm{SN}^\mathrm{disk}=320$~kms$^{-1}$, $V_\mathrm{SN}^\mathrm{burst}=320$~kms$^{-1}$, and $\gamma_\mathrm{SN}=3.2$. The form of Equation~\ref{eq:loading} is motivated by the argument that the efficacy of mass ejection from the galaxy should vary inversely on the depth of the gravitational potential well, which is characterised here by $V_c$. 

Motivated by \cite{Font2011} and \cite{Hou2016}, we introduce an additional parameter, $V_\mathrm{sat}$, which acts as a threshold circular velocity below which the supernova feedback saturates. More specifically, we set $\gamma_\mathrm{SN}=0$ when $V_c < V_\mathrm{sat}$. Such a parameter sets an upper limit to the mass loading factor and prevents it from becoming excessively large at lower masses. $V_\mathrm{sat}$ is treated as an adjustable parameter, which is not included in the default \galform{} model (in effect, $V_\mathrm{sat}=0$). 

Once gas has been ejected from the galaxy due to supernova feedback, it is assumed to accumulate in a reservoir beyond the virial radius of the halo. This gas may eventually return to the hot gas reservoir of the halo at a rate given by:
\begin{equation} \label{eq:return}
    \dot{M}_\mathrm{return} = \alpha_\mathrm{ret} \frac{M_\mathrm{res}}{\tau_\mathrm{dyn,halo}},
\end{equation}
where $M_\mathrm{res}$ is the mass of the gas reservoir, $\tau_\mathrm{dyn,halo}$ is the halo dynamical time, and $\alpha_\mathrm{ret}$ is an adjustable parameter, set to $0.64$ in the fiducial model. 

\subsubsection{Reionisation}
\label{sec:reion}

In \galform{}, we impose reionisation using a simple two-parameter model. In short, this parametrisation mimics the effect of ionisation from a global UV background by turning off gas cooling in a halo with circular velocity, $V_c$, if $V_c<V_\mathrm{crit}$ at $z<z_\mathrm{reion}$. Here, $V_\mathrm{crit}$ is the so-called `filtering scale' for reionisation, and represents the minimum virial velocity (temperature) of a halo that is able to cool gas after the IGM is heated during the reionisation process \citep[see, e.g.,][]{Llambay2020}. $z_\mathrm{reion}$ is the redshift at which reionisation occurs; both are adjustable parameters. In the fiducial model, we set $V_\mathrm{crit}=30$ kms$^{-1}$ \citep[c.f.][]{Gnedin2000,Thoul1996,Okamoto2008} and $z_\mathrm{reion}=6$. Clearly, this model is an oversimplification, which treats reionisation as a homogeneous and instantaneous process. This is not a limitation for our present investigation in which we are  interested in quantifying how different choices of parameters like these affect the properties of Milky Way satellites in a systematic way. Furthermore, it turns out that this two parameter model represents a reasonable approximation to a comprehensive, self-consistent calculation of reionisation in \galform{} performed by \cite{Benson2002b}. \cite{Font2011} also showed that this `shortcut' remains a good approximation even when local ionising sources are included in addition to the global ionising background.

\begin{table*}
    \centering
    \begin{tabularx}{\textwidth}{ l  X  c  c  c }
    \toprule
      {\bf Parameter} & {\bf Description} & {\bf Range} & {\bf Fiducial value} & {\bf Reference/Equation}  \\
    \toprule
      \multicolumn{4}{l}{\bf Cosmology} \\
      $\Omega_{\mathrm{m}}$ & \rdelim\}{4}{*}[ WMAP-7] & $-$ & 0.272 & \rdelim\}{4}{*}[ \citet{Komatsu2011}] \\
      $\Omega_\Lambda$ & & $-$ & 0.728 \\
      $\sigma_8$ & & $-$ & 0.816 \\
      $h$ & & $-$ & 0.704 \\
    \midrule
      \multicolumn{4}{l}{\bf Star formation in disks} \\
      $\nu_{\mathrm{SF}}$ & molecular gas depletion timescale for star formation in disks & $\left[0.2-1.7\right]$ Gyr$^{-1}$ & $0.74$ Gyr$^{-1}$ & Equation~\ref{eq:sf_disk} \\
    \midrule
      \multicolumn{4}{l}{\bf Starburst triggering in galaxy mergers and instabilities} \\
      $f_\mathrm{burst}$ & threshold on mass ratio for starbursts & $0.01-0.2$ & $0.05$ & Section~\ref{sec:starform} \\
      $f_\mathrm{ellip}$ & threshold on mass ratio for major mergers & $0.2-0.5$  & $0.3$ & Section~\ref{sec:starform} \\
      $F_{\mathrm{stab}}$ & threshold for disk instability & $0.4-1.1$ & 0.9 & Equation~\ref{eq:f_disk} \\
    \midrule
      \multicolumn{4}{l}{\bf Supernova feedback} \\
      $V_{\mathrm{SN}}^{\mathrm{disk}}$ & normalisation in mass loading factor for disk star formation & $\left[5-400\right]$ kms$^{-1}$ & $320$ kms$^{-1}$ & \rdelim\}{3}{*}[ Equation~\ref{eq:loading}] \\
      $V_{\mathrm{SN}}^{\mathrm{burst}}$ & normalisation in mass loading factor in starbursts & $\left[5-400\right]$ kms$^{-1}$ & $320$ kms$^{-1}$ \\
      $\gamma_{\mathrm{SN}}$ & power-law index for mass loading factor (velocity scaling) & $0-5$ & 3.2 \\
      $\alpha_{\mathrm{ret}}$ & multiplier for hot gas reincorporation timescale & $0-4$ & 0.64 & Equation~\ref{eq:return} \\
      $V_{\mathrm{sat}}$ & threshold circular velocity below which feedback is saturated & $\left[0-80\right]$ kms$^{-1}$ & $-$ & Section~\ref{sec:SNfb} \\
    \midrule
      \multicolumn{4}{l}{\bf Reionisation} \\
      $z_{\mathrm{reion}}$ & redshift of reionisation & $2-15$ & 6 & \rdelim\}{2}{*}[ Section~\ref{sec:reion}] \\
      $V_\mathrm{crit}$ & critical filtering scale for reionisation & $\left[5-60\right]$ kms$^{-1}$ & $30$ kms$^{-1}$ \\
    \bottomrule
    \end{tabularx}
    \caption{A summary of the most important parameters (first column) and their physical interpretation (second column) that are  relevant to this work. The range over which each parameter is varied is shown in the third column, whereas the fourth column lists its fiducial value in the \citet{Lacey2016} model. Finally, the last column provides a reference to the relevant section or equation in which the parameter is first introduced. Note that \citet{Lacey2016} assume $z_\mathrm{reion}=10$, whereas our default choice is $z_\mathrm{reion}=6$ based on the work of \citet{Bose2018}. The cosmological parameters are listed for completeness; their values are kept fixed at the WMAP-7 values throughout the entirety of this paper.}
    \label{tab:params}
\end{table*}

\subsubsection{The evolution of mass and metals}
\label{sec:chemistry}

With the main physical processes in hand, we are now able to describe the system of equations that govern the exchange of mass and metals between the different baryonic components associated with a halo. These are: the hot gas in haloes ($M_\mathrm{hot}$), the cold gas in galaxies ($M_\mathrm{cold}$), the reservoir of ejected gas {\it outside} haloes ($M_\mathrm{res}$), and the stellar mass in galaxies ($M_\star$). Their evolution is given by:
\begin{align} \label{eq:evolution}
    \dot{M}_\mathrm{hot} &= -\dot{M}_\mathrm{acc} + \alpha_\mathrm{ret}\frac{M_\mathrm{res}}{\tau_\mathrm{dyn,halo}} \\
    \dot{M}_\mathrm{cold} &= \dot{M}_\mathrm{acc} - \left(1-R+\beta\right)\psi \\
    \dot{M}_\star &= \left(1-R\right)\psi \\
    \dot{M}_\mathrm{res} &= \beta \psi - \alpha_\mathrm{ret}\frac{M_\mathrm{res}}{\tau_\mathrm{dyn,halo}},
\end{align} 
in which we have made use of the instantaneous recycling approximation \citep{Tinsley1980}, such that there is no time delay between when stars form and when they die and eject gas and metals. In this case, the rate of gas ejection by dying stars into the
cold component is $R\psi$, where $R$ is the so-called `returned fraction', and is specified by the IMF assumed. 

The evolution of metals is governed by a very similar set of equations:
\begin{align} \label{eq:metals}
    \dot{M}_\mathrm{hot}^Z &= -Z_\mathrm{hot}\dot{M}_\mathrm{acc} + \alpha_\mathrm{ret}\frac{M_\mathrm{res}^Z}{\tau_\mathrm{dyn,halo}} \\
    \dot{M}_\mathrm{cold}^Z &= Z_\mathrm{hot}\dot{M}_\mathrm{acc} + \left[p - \left(1-R+\beta\right)Z_\mathrm{cold}\right]\psi \label{eq:yield} \\
    \dot{M}_\star^Z &= \left(1-R\right)Z_\mathrm{cold}\psi \\
    \dot{M}_\mathrm{res}^Z &= \beta Z_\mathrm{cold} \psi - \alpha_\mathrm{ret}\frac{M_\mathrm{res}^Z}{\tau_\mathrm{dyn,halo}},
\end{align} 
where we have defined $M_\mathrm{hot}^Z$ as the mass of metals in the hot component, and $Z_\mathrm{hot}=M_\mathrm{hot}^Z/M_\mathrm{hot}$ as the corresponding metallicity (and similarly for the remaining components). The term $p\psi$ quantifies the ejection rate of newly formed metals into the interstellar medium (ISM) by dying stars, in which $p$ is the yield, and which also depends on the choice of IMF. \galform{} assumes two different IMFs: a \cite{Kennicutt1983} IMF for quiescent star formation in disks ($R=0.44$ and $p=0.021$) and a top-heavy IMF in starbursts ($R=0.54$ and $p=0.048$). Again, the instantaneous recycling approximation has been assumed for the evolution of metals.

It is worth emphasising an important point at this stage. While some of the physical prescriptions incorporated in \galform{} are informed by the properties of the dwarf galaxy population \citep[in, for e.g., the legacy developments made by][for the physics of reionisation]{Benson2002a,Font2011}, the specific parameter set used to {\it calibrate} any instance of \galform{} \citep[such as in][]{Lacey2016} never uses the properties of the Milky Way satellite population.

Table~\ref{tab:params} summarises the list of parameters varied in this work, and includes their descriptions, fiducial values, and the range over which they are varied. In total, we end up with a list of 11 adjustable parameters, which represent a range of physical processes in the model. In the following section, we will describe how their influence on the satellite population can be quantified in a systematic way. 

\subsection{Variance-based sensitivity analysis using Sobol' indices}
\label{sec:sobol}

To understand how changes in our input parameters contribute to the variance in the satellite galaxy properties predicted by \galform{} (specifically, the luminosity function and the mass-metallicity relation), we employ a global sensitivity analysis. In particular, we use the Sobol' method \citep{Sobol2001}, which decomposes the variance of the model output into contributions that can be attributed to individual input parameters or sets of parameters.

Let our model be represented by a function $Y = f(\mathbf{X})$, where $Y$ is a scalar output (e.g., the predicted abundance of Milky Way satellites in some bin of absolute V-band magnitude) and $\mathbf{X} = (X_1, X_2, \dots, X_k)$ is a vector of $k$ input parameters (i.e., those listed in Table~\ref{tab:params}), each with a defined probability distribution. The conditional expectation $E(Y|X_i=x_i^*)$ is the expected value of the output $Y$ when the input parameter $X_i$ is held fixed at a value $x_i^*$. Formally, we define this as:
\begin{equation} \label{eq:expectation}
E(Y|X_i=x_i^*) = \int_{X_{\sim i}} f(x_1, \dots, x_i^*, \dots, x_k) \, p(x_{\sim i}) \, dx_{\sim i},
\end{equation}
where $X_{\sim i}$ denotes the set of all input parameters except for $X_i$ and $p(x_{\sim i})$ is the joint probability density function of all parameters except $X_i$. 

The total variance, $\mathrm{Var}(Y)$, of the output can be decomposed as:
\begin{equation} \label{eq:variance}
\mathrm{Var}(Y) = \sum_{i=1}^{k} \mathrm{Var}_i + \sum_{i<j}^{k} \mathrm{Var}_{ij} + \dots + \mathrm{Var}_{12\dots k},
\end{equation}
where $\mathrm{Var}_i = \mathrm{Var}(E(Y|X_i))$ is the variance of the conditional expectation of the output given that $X_i$ is fixed. This is the first-order effect of $X_i$ on $Y$. The higher-order terms $\mathrm{Var}_{ij}$, etc., represent all possible interactions {\it between} parameters.

Armed with these definitions, the Sobol' indices can then be defined as dimensionless quantities obtained by dividing the variance terms by the total variance. In this paper, we will consider two definitions in particular:
\begin{enumerate}
    \item The normalised {\it first-order} index, $S_i$, which measures the primary effect of the parameter $X_i$ on the output variance:
    \begin{equation} \label{eq:Si_eq}
        S_i = \frac{\mathrm{Var}(E(Y|X_i))}{\mathrm{Var}(Y)}, 
    \end{equation}
    wherein a high $S_i$ indicates that parameter $X_i$ is influential on its own. Defined in this way, if $S_i=1$, the entire variance in $Y$ is attributable to parameter $X_i$; conversely, if $S_i=0$, then $Y$ is independent of $X_i$.
    \item The {\it total} sensitivity index, $S_{\mathrm{T}i}$, which measures the contribution of $X_i$ including all (i.e., second- and higher-order) its interactions with other parameters:
    \begin{equation} \label{eq:ST_eq}
        S_{\mathrm{T}i} = \frac{E(\mathrm{Var}(Y|X_{\sim i}))}{\mathrm{Var}(Y)} = 1 - \frac{\mathrm{Var}(E(Y|X_{\sim i}))}{\mathrm{Var}(Y)}.
    \end{equation}
\end{enumerate}
The total sensitivity index is a unique feature of variance-based sensitivity analysis, enabling us to capture complex interactions between parameters when predicting some output statistic. This is one of the primary motivators for the use of this method in this investigation. 

\begin{figure*}
    \centering
    \includegraphics[width=0.49\linewidth]{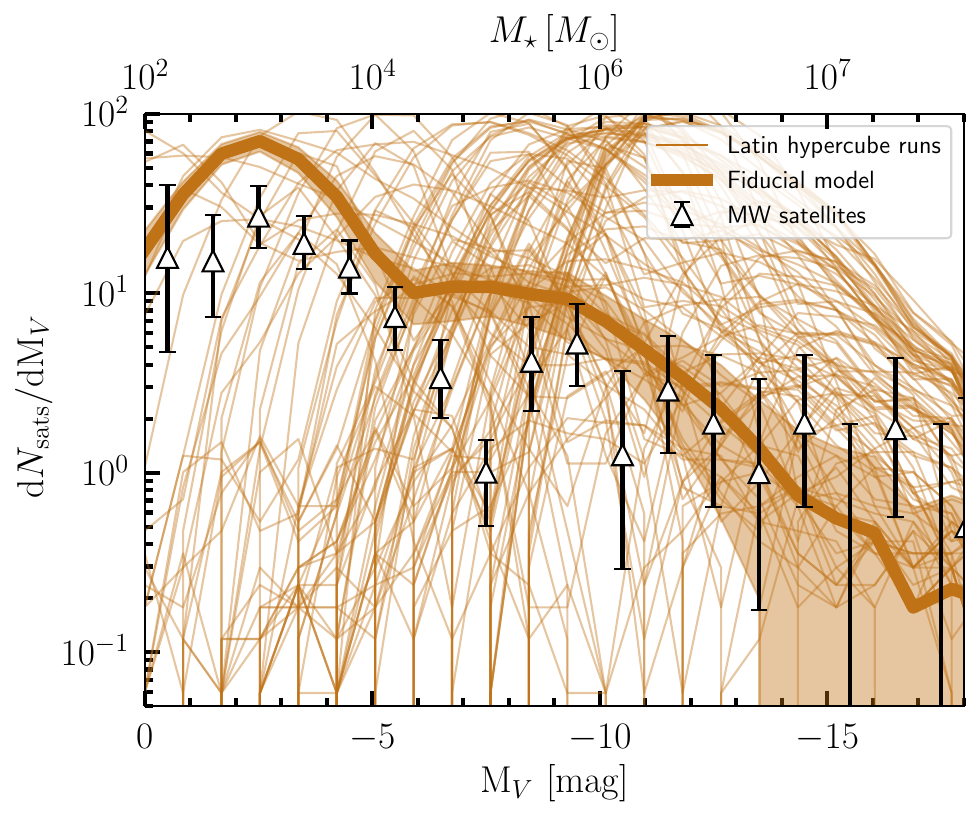}
    \includegraphics[width=0.49\linewidth]{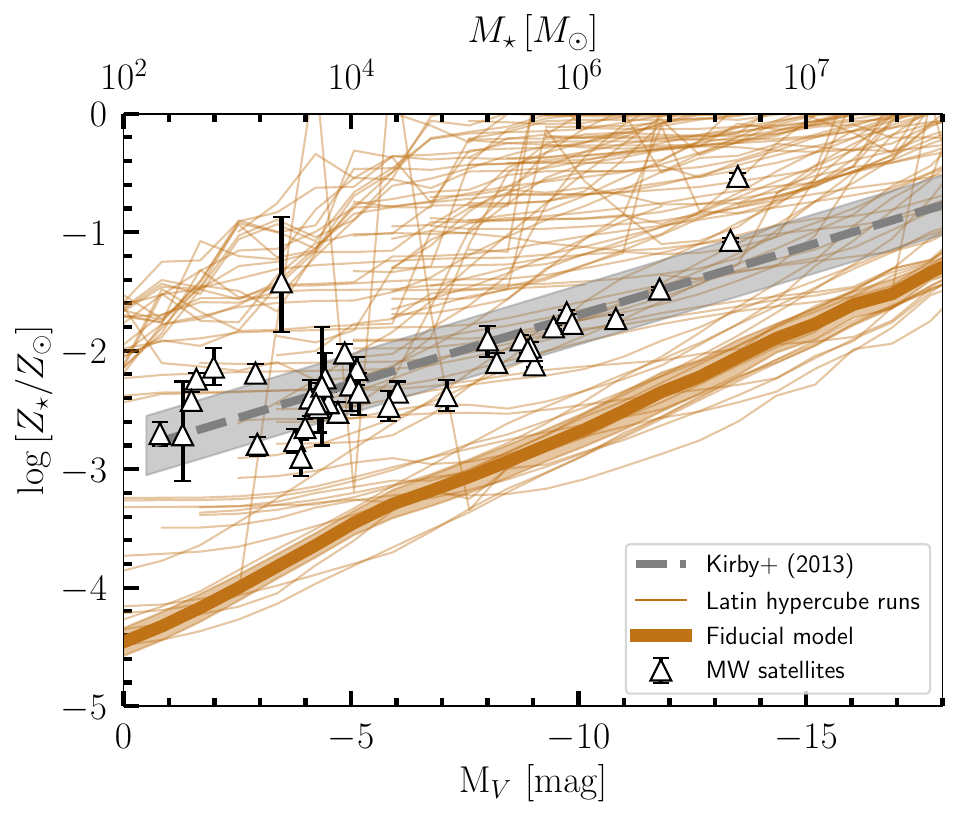}
    \caption{A comparison of satellite properties predicted by \galform{} and observational data for the Milky Way (MW). In both panels, the results for 100 examples extracted from the Latin hypercube parameter space exploration are shown as thin copper lines, while the fiducial model is indicated by the thick solid curve of the same colour. These 100 runs were selected randomly from the full sample of 2600 runs; we find that these largely encompass the full range of predictions made by the model. {\bf Left panel}: the satellite luminosity function, which shows the number of satellites as a function of absolute V-band magnitude, $\mathrm{M}_V$. The model predictions are compared against the combined satellite count for the Milky Way and M31, as described in Section~\ref{sec:emulator}; these are shown as black triangles with error bars -- hereafter, we label this combined data set ``MW satellites'' for simplicity. The fiducial \galform{} model provides a reasonable qualitative agreement with the observed satellite counts, particularly for brighter satellites ($\mathrm{M}_V < -10$). {\bf Right panel}: the satellite mass-metallicity relation, with stellar metallicity ($Z_\star$) shown as a function of $\mathrm{M}_V$. \galform{} predictions are compared with data compiled by \citet{Simon2019} for Milky Way satellites and the best-fit relation from \citet{Kirby2013} (dashed line, with the associated shaded region showing a 0.25 dex uncertainty around the median relation). The fiducial model predicts systematically lower metallicities at a fixed magnitude than what is observed in the data. In both panels, we find that \galform{} models span a range of predictions, including many examples that are likely implausible and in strong tension with the data. The thin lines are matched across the left- and right-hand panels, and the shaded regions mark the 68\% scatter around the median in the fiducial run.}.
    \label{fig:all_models}
\end{figure*}

At the crux of any kind of sensitivity analysis is the assumption that the input parameters of the model are independent. In other words, the sensitivity analysis method does not allow us to recognise parameter combinations that have intrinsic correlations, or which represent unphysical combinations from the perspective of a theory of galaxy formation. While this limits the applicability of this method to more complex model architectures, these limitations do not apply to semi-analytic models like \galform{}, in which model parameters can be varied independently. This inherent feature of the semi-analytic method, therefore, allows us to systematically experiment with choices made in the model and to examine how these choices impact the statistics predicted by it. The use of Sobol' indices in the context of galaxy formation models was first presented in \cite{Oleskiewicz2020} and \cite{Elliott2021}. An exploration of variance-based sensitivity analysis in the context of Milky Way satellites was presented in \cite{Gomez2014}, albeit with a very different model infrastructure to the one presented here.

Prior to computing Sobol' indices from our model predictions, it is important to define a dense, well-sampled  parameter space that covers the 11-dimensional manifold defined by the \galform{} model parameters listed in Table~\ref{tab:params}. To do this in an unbiased and optimised manner, we employ the use of a Saltelli sampler \citep{Saltelli2002}. This allows us to predict the satellite luminosity function and mass-metallicity relation over a large number of points in the parameter space; the net result is that Equations~\ref{eq:expectation}-\ref{eq:ST_eq} can now be approximated using a Monte Carlo approach. For further details of Saltelli sampling, we refer the reader to the discussion in Appendix~\ref{sec:saltelli}. We use the \texttt{SALib}\footnote{\url{https://salib.readthedocs.io}} library \citep{Herman2017,Iwanaga2022} to perform the Saltelli sampling and to compute Sobol' indices.

\subsection{A neural network emulator for the \galform{} satellite population}
\label{sec:emulator}

As the focus of this paper is to investigate the satellite population of the Milky Way, we limit ourselves to only consider the assembly history of Milky Way-mass haloes as a first step. We generate 100 realisations of a Milky Way-like halo, defined as a host halo with a present-day mass of $10^{12}\,\mathrm{M}_\odot$; by using multiple realisations, we are able to measure not just the median, but also the {\it scatter} predicted for any given statistic. The merger trees are built to resolve haloes as small as $M_\mathrm{halo}\sim10^6\,\mathrm{M}_\odot$, which is well below the mass scale in which the smallest galaxies form in our model (which are haloes with virial velocities $\sim16-17$ kms$^{-1}$, corresponding to the atomic hydrogen cooling limit; see Appendix~\ref{sec:gal_host}).

Each \galform{} run takes around 3 hours on a single core using these settings. While this is not prohibitive, it is not ideal given the enormous number of samples needed for the evaluation of Sobol' indices\footnote{Our \galform{} emulator is evaluated at more than 49,000 points in the 11-dimensional parameter space to obtain the results presented in Section~\ref{sec:main}.}. For this reason, we design a neural network emulator, trained on the outputs of full \galform{} simulations, to expedite this process. Our approach builds upon early work by \cite{Bower2010} who used a Bayesian emulator to explore the parameter space of the \galform{} model, albeit focusing on the luminosity function of field galaxies.

We begin by performing a series of \galform{} runs, sampling 2600 points in the 11-dimensional input space in which each parameter is varied over the ranges shown in Table~\ref{tab:params}. This input space is sampled using a Latin hypercube. From each of these runs, we compute the (binned) luminosity function and the mass-metallicity relation of the associated satellite galaxy populations. As each host halo merger tree is generated 100 times (using different random seeds) for each of the 2600 \galform{} runs, we are able to measure both the median relation as well as the scatter for each set of parameter choices.

Figure~\ref{fig:all_models} shows the predictions for 100 of these runs, which have been selected at random. The panel on the left shows the satellite luminosity function, whereas the panel on the right shows the mass-metallicity relation for the same runs. In each panel, the lower x-axis shows the (present-day) absolute V-band magnitude, whereas the upper x-axis denotes the corresponding stellar mass. Note that while the satellite luminosity function is typically shown as a cumulative quantity, we opt in this paper to show its differential version. The thick copper line in each panel shows the prediction of the fiducial model (i.e., based on the parameters listed in the fourth column of Table~\ref{tab:params}); the thinner lines are predictions from other parameter variations. In both panels, observational data are shown using black triangles with error bars. For the satellite number counts, we have combined the luminosity function for the Milky Way's satellite population inferred by \cite{Newton2018} with the corresponding data for M31 compiled from the PAndAS survey \citep{McConnachie2009,McConnachie2012,Ibata2014,Martin2016} using the procedure described in \citet[][their Section 3]{Bose2018}. This is done to boost the statistical fidelity of the data we compare to. Finally, observational measurements of the stellar abundances of Milky Way satellites are obtained from Table 1 in \cite{Simon2019}; the scaling relation derived by \cite{Kirby2013} is also shown for reference.

A couple of interesting features can be noticed immediately. We find that the fiducial model does a reasonable job of reproducing the satellite luminosity function, not just in terms of the overall number counts (particularly at the bright end), but also in its shape (especially considering that \galform{} has not been calibrated in this regime whatsoever). In particular, the luminosity shows a distinctive shape in which there is a `kink' at $M_\mathrm{V}\approx-5$ ($M_\star\approx10^4\,\mathrm{M}_\odot$). This feature was first identified in \cite{Bose2018}, and is the characteristic scale that separates the galaxies that form prior to reionisation from those that assemble after; its location can be mapped directly to the parameter $V_\mathrm{crit}$. The model somewhat overpredicts the number of ultrafaints, but this can be accommodated through a recalibration of parameters, or by considering a lower host halo mass. 

The mass-metallicity relation, shown in the right-hand panel of Figure~\ref{fig:all_models}, paints a very different picture. Here, we find that the fiducial model vastly underpredicts the stellar metallicities of galaxies, with the discrepancy growing with decreasing stellar mass. For example, at $M_\star\sim10^7\,\mathrm{M}_\odot$, the model predicts metallicities that are a factor of 3 lower than the data; at $M_\star\sim10^3\,\mathrm{M}_\odot$, this increases to a factor of 30. The fact that the mass-metallicity relation is not reproduced in conjunction with the luminosity function (under standard implementations of feedback and reionisation) in \galform{} has previously been noted by \cite{Font2011}, as well as in the semi-analytic model of \cite{Lu2017}. This is by no means a shortcoming that is unique to semi-analytic models; indeed, this is also seen in hydrodynamical simulations of dwarf galaxies \citep[e.g.,][]{Munshi2019,CWheeler2019,Agertz2020,Sanati2023,Azartash2024,CWheeler2025}. With that being said, we note that this underprediction is not universal; simulations such as EAGLE \citep{Schaye2015} and Auriga \citep{Grand2017}, as well as other semi-analytic models \citep{DeLucia2014}, predict higher metallicities that are in better agreement with, or even overestimate, the data. The recent COLIBRE simulations \citep{Schaye2025} also demonstrate rather good agreement with the observed data for this relation. The fiducial \galform{} model predicts a relation with a relatively constant slope across the entire mass range.

\begin{figure}
    \centering
    \includegraphics[width=\columnwidth]{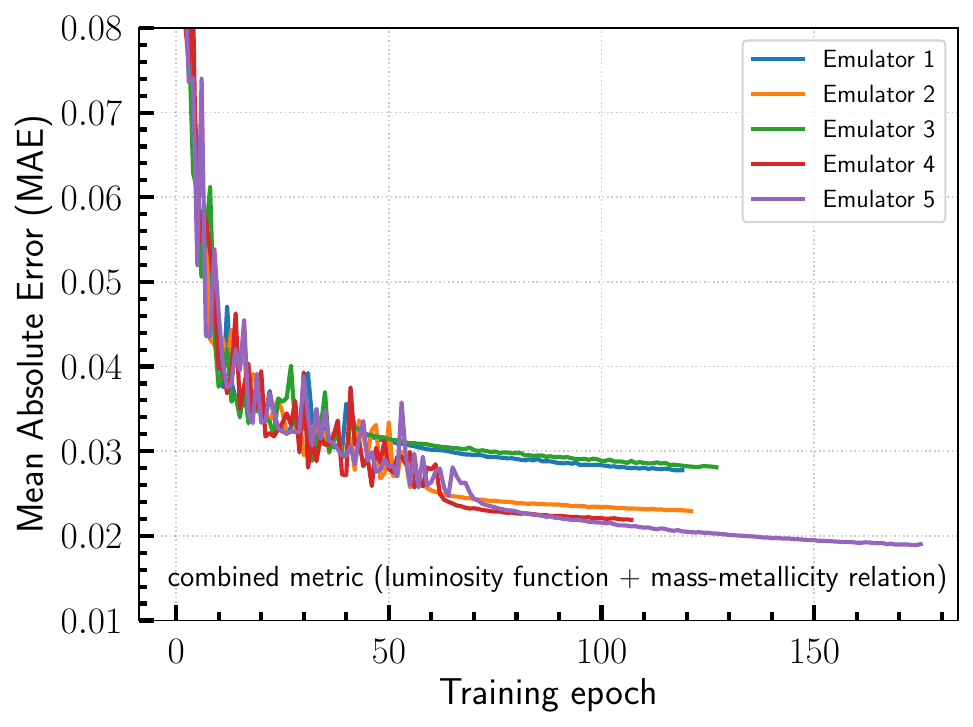}
    \caption{The validation loss as a function of training epoch for five distinct emulator models. In particular, we show the Mean Absolute Error (MAE), a combined metric that averages the loss from both the luminosity function and the mass-metallicity relation in some weighted combination. Each coloured line corresponds to an identical model architecture trained on a different fold of the training data in a 5-fold cross-validation scheme. We find that the validation error for all models decreases rapidly during the initial 50 epochs of training, after which performance plateaus as the models converge. All five models achieve a similar final MAE, reaching a value between 0.02 and 0.03 after $\sim$120 epochs.}
    \label{fig:combined_MAE}
\end{figure}

Turning our attention now to the predictions of the parameter variations, we find that there is quite a range of possibilities that can be accommodated by the \galform{} model. Clearly, many of these are highly implausible, where, in some cases, the number counts in the faintest luminosity bins are nearly non-existent, producing a luminosity function that disagrees with the data completely. The models also span the full range of possibilities in terms of galaxy metallicities, ranging from examples that are qualitatively consistent with the data, to those that overpredict the abundances by orders of magnitude. It is also interesting to note that some models predict a flattening in the relation at lower stellar masses ($M_\star\lesssim10^4\,\mathrm{M}_\odot$), where the metallicities appear to plateau, and for which there may be some tentative evidence in the data \citep[see, e.g.,][]{Fu2023}.

With the input data now defined, we are in a position to train our neural network emulator following a strategy that is similar to the work of \cite{Elliott2021} and \cite{Madar2024}. The 2600 \galform{} runs are first partitioned to hold out a dedicated test sample (100 runs), the remainder being split into training (80\%) and validation (20\%) sets. We use the \texttt{TensorFlow}\footnote{\url{https://www.tensorflow.org/}} package \citep{Tensorflow2015} to construct a network architecture consisting of five shared hidden layers, each with 512 neurons and a Leaky Rectified Linear Unit \citep[Leaky ReLU,][]{Maas2013} activation function. This shared core feeds into six distinct output layers that simultaneously predict the median and scatter of the satellite luminosity function and the mass-metallicity relation. To train the network, we minimise a loss function defined by the Mean Absolute Error (MAE) between the predictions of the emulator and the true \galform{} outputs. We train an ensemble of five independent models to enhance the stability of the predictions made by the emulator. The final output is the mean prediction across the ensemble \citep[the advantages of ensemble averaging emulator predictions is described in detail in][]{Madar2024}. The training process for each model proceeds in two stages: an initial phase using the Adaptive Momentum Estimation (Adam) optimiser \citep{Kingma2014} with a learning rate of $10^{-3}$, followed by a fine-tuning phase using the RMSprop optimiser \citep{Tieleman2012} with a learning rate of $10^{-5}$. We employ a custom early stopping condition that halts training when a weighted combination of the validation MAE across the luminosity function and the mass-metallicity relation no longer improves over 30 epochs. This avoids overfitting, and optimises the overall performance of the model.

\begin{figure*}
    \centering
    \includegraphics[width=\textwidth]{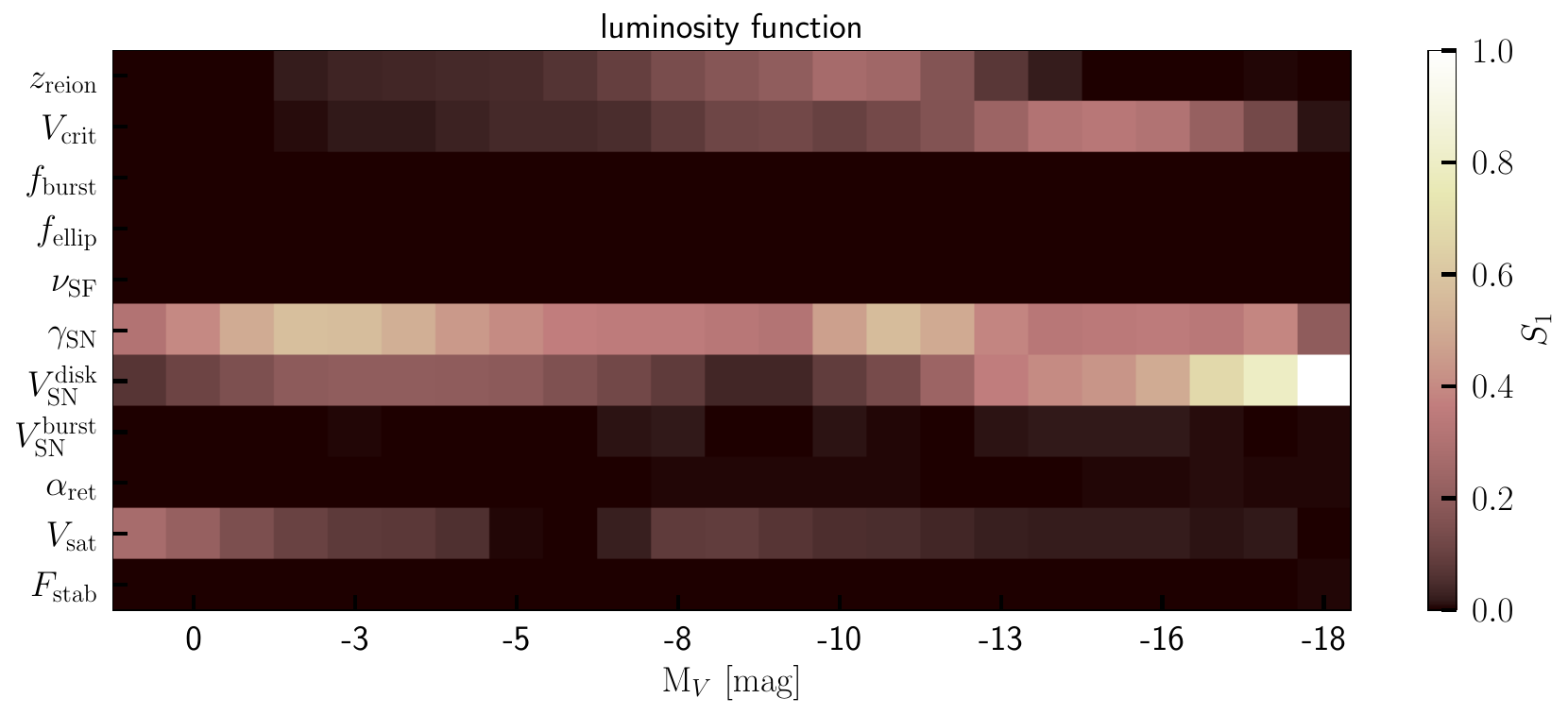}
    \caption{First-order Sobol' sensitivity indices, $S_1$, for the $z=0$ satellite luminosity function. Each pixel represents the sensitivity of the \galform{} predictions in a bin of absolute V-band magnitude, $M_V$, to variations in each of the 11 input parameters of the model (see Section~\ref{sec:galform} for their descriptions). The value of $S_1$, indicated by the adjoining colour scale, quantifies the fraction of the output variance in the luminosity function that is attributable directly to a single parameter defining the model. Higher values of $S_1$ correspond to parameters that are highly influential in the predicted abundance of satellite galaxies, while values close to zero represent parameters that have little to no influence. We find that the set of influential parameters changes as a function of satellite luminosity. The abundance of bright satellites ($M_V \lesssim -13$) is predominantly sensitive to parameters governing supernova feedback, specifically the parameters $\gamma_\mathrm{SN}$ and $V_{\mathrm{SN}}^{\mathrm{disk}}$, which define the mass loading factor. For fainter satellites, the sensitivity to the redshift and filtering scale for reionisation ($z_{\mathrm{reion}}$ and $V_\mathrm{crit}$, respectively) increases significantly. Parameters relating to mergers, quiescent star formation in disks and starbursts have no discernible impact on the final outcome.}
    \label{fig:S1_SLF}
\end{figure*}

Figure~\ref{fig:combined_MAE} shows the validation loss of this combined metric as a function of training epoch. The coloured lines represent separate instances of our neural network, which can each be thought of as a different initialisation of weights prior to training. In each case, we observe that the MAE loss function declines rapidly within the 50 first epochs of training. The onset of the fine-tuning stage is recognised by the sudden drop in the loss function, at which point the learning rate is reduced by a factor of a hundred. Note that at this point the curves also become smoother, eventually asymptoting to some minimum MAE before the convergence criterion is met. We have experimented with several different combinations for the number of hidden layers, numbers of neurons per layer, choice of activation function etc., and find that the model architecture we have presented here offers an acceptable compromise between model accuracy and training time. The hold-out set of `unseen' data is used at the end of the training and validation process to analyse the performance of the (ensemble-averaged) neural network predictions. The training process takes around two minutes per model instance on GPUs; once trained, the luminosity function and mass-metallicity relations for tens of thousands of \galform{} parameter combinations can be made in mere seconds.

\section{A sensitivity analysis of the Milky Way satellite population}
\label{sec:main}

Having constructed an emulator that is able to predict the luminosity function and mass-metallicity relation for \galform{} galaxies accurately and efficiently, we are now in a position to quantify their sensitivity to the model parameters using the formalism described in Section~\ref{sec:sobol}.

\subsection{The first-order influence of model parameters}
\label{sec:sobol_S1}

Figure~\ref{fig:S1_SLF} shows the results of our sensitivity analysis for the satellite luminosity function. The figure displays the first-order Sobol' index, $S_1$, which quantifies the fractional contribution of each of the 11 input parameters (listed along the vertical axis) to the total variance in satellite counts within bins of absolute V-band magnitude, $\mathrm{M}_V$. In essence, this is another way to visualise the satellite luminosity function, with each bin now representing the {\it influence of individual parameters} to the satellite number counts. 

For the brighter satellites, with $\mathrm{M}_V \lesssim -13$, the luminosity function is overwhelmingly sensitive to the parameters governing supernova feedback. In this regime, the power-law index of the mass loading factor, $\gamma_\mathrm{SN}$, and the feedback normalisation in galactic disks, $V_\mathrm{SN}^{\mathrm{disk}}$, are the dominant sources of variance, as indicated by their high $S_1$ values. This suggests, perhaps unsurprisingly, that the number of bright satellites is regulated primarily by the efficiency of stellar feedback in quenching star formation within their host dark matter haloes. Other physical processes, such as reionisation, have a negligible impact on the abundance of these brighter satellites.

\begin{figure*}
    \centering
    \includegraphics[width=\textwidth]{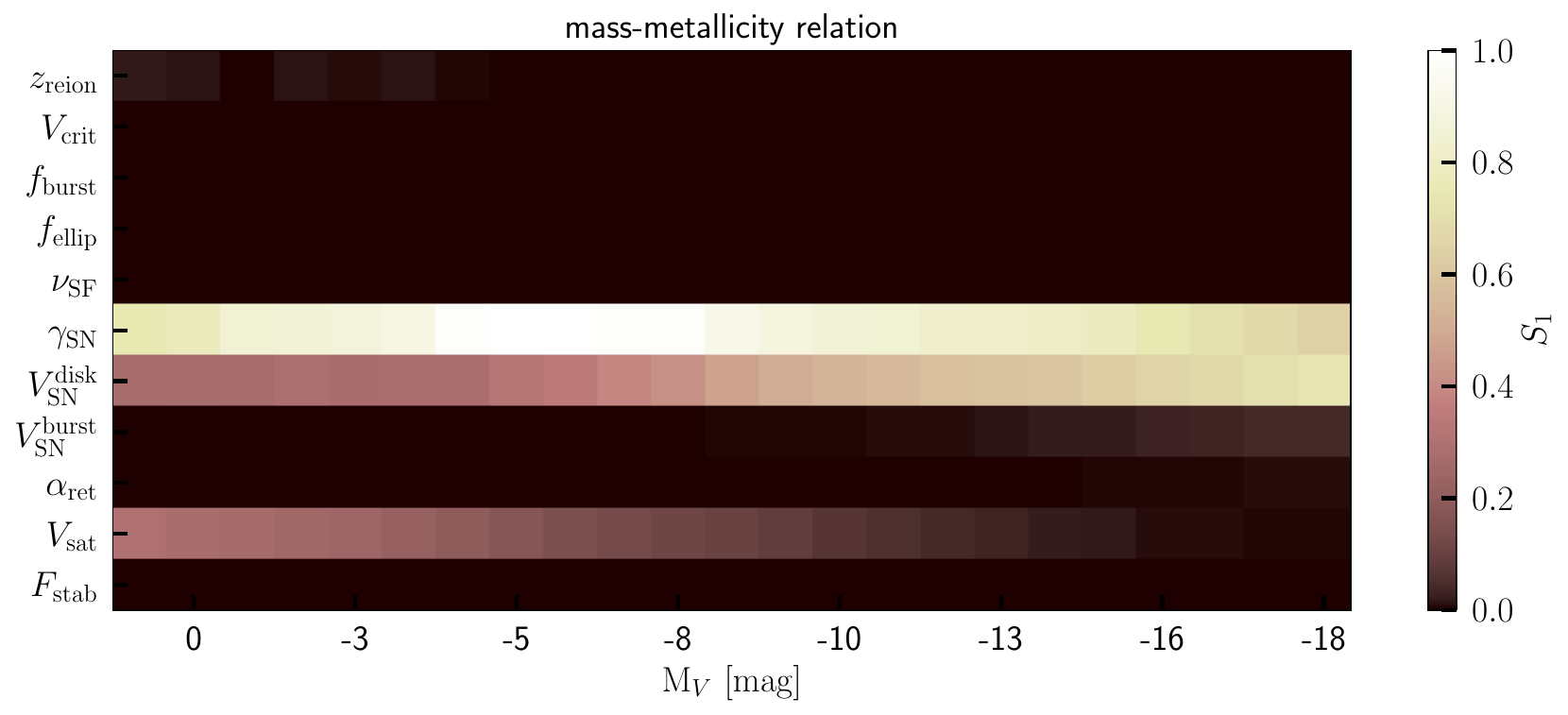}
    \caption{As Figure~\ref{fig:S1_SLF}, but now showing the case of the satellite mass (luminosity)-metallicity relation.}
    \label{fig:S1_SMZ}
\end{figure*}

As we move towards fainter galaxies, the physical picture becomes more complicated. The set of influential parameters expands as we move from the bright to the faint end of the luminosity function, rather than a simple switch in the dominance of a given process from one regime to the other. The power-law scaling of the mass loading factor, $\gamma_\mathrm{SN}$, remains highly influential across all magnitude bins, indicating that supernova feedback is a crucial physical process for all satellites. However, for these fainter systems, other parameters rise to a comparable level of importance. Notably, the sensitivity to the redshift of reionisation, $z_\mathrm{reion}$, increases dramatically, as does that of the  filtering scale, $V_\mathrm{crit}$. Increasing (decreasing) this parameter means that higher (lower) mass haloes are now quenched by reionisation. As we show in Figure~\ref{fig:mstar-mhalo}, the range over which this parameter is varied ($5\lesssim V_\mathrm{peak}/\mathrm{kms}^{-1}\lesssim60$) impacts satellites over a large range in V-band magnitude, including satellites as bright as $\mathrm{M}_V\approx-15$ mag ($M_\star\approx10^7\,\mathrm{M}_\odot$) even in the fiducial model. Similarly, the parameter $V_\mathrm{sat}$, which sets a ceiling on the mass loading factor, becomes a key parameter in this regime. We observe a gap in the influence of $V_\mathrm{sat}$ around $\mathrm{M}_V \approx -5$. This likely marks the transition scale where the suppression of galaxy formation shifts from being dominated by reionisation (governed by $V_\mathrm{crit}$) to being dominated by the saturation in supernova feedback. We note, however, that its effect is subdominant to the parameters we have described above. 

We turn our attention next to Figure~\ref{fig:S1_SMZ}, which shows the first-order Sobol' indices for the mass-metallicity relation for satellite galaxies. All visual elements of this figure (as well as their interpretations) are identical to those in Figure~\ref{fig:S1_SLF}. A clear difference from what was observed for the satellite luminosity function now emerges. Although the supernova mass loading parameters ($\gamma_\mathrm{SN}$ and $V_\mathrm{SN}^\mathrm{disk}$) are still the most influential parameters, their status as the dominant physical processes is elevated in the case of the chemical abundances of \galform{} satellites, which is inferred from the larger $S_1$ values in each magnitude bin. Interestingly, we note that while the influence of the normalisation of the mass loading in disks ($V_\mathrm{SN}^\mathrm{disk}$) increases as we move towards brighter satellites (as it was in the case of the luminosity function), the power-law index of this scaling ($\gamma_\mathrm{SN}$) is equally influential in all magnitude bins. We also note that the parameter that determines the scale at which feedback efficiency saturates ($V_\mathrm{sat}$) is of comparable importance to $V_\mathrm{SN}^\mathrm{disk}$ in the regime $\mathrm{M}_V\gtrsim-6$ mag. In stark contrast to the satellite luminosity function, however, we find that reionisation has no effect on the mass-metallicity relation.

Finally, other parameters relating to the efficiency of quiescent star formation ($\nu_\mathrm{SF}$), feedback strength in starbursts ($V_\mathrm{SN}^\mathrm{burst}$), the triggering of merger- and disk instability-induced starbursts ($f_\mathrm{burst}$, $f_\mathrm{ellip}$, $F_\mathrm{stab}$) and the hot gas reincorporation timescale ($\alpha_\mathrm{ret}$) have essentially no impact on either the satellite luminosity function or the mass-metallicity relation, at least within the ranges explored in this paper. This is in contrast to the conclusions of \cite{Bower2010}, who found $F_\mathrm{stab}$ to be influential for the break of the field galaxy luminosity function. This is likely because the formation of instabilities is less critical for the evolution of the lower mass satellites considered here. A similar conclusion is also reached by \cite{Oleskiewicz2020}, who note the descresing influence of this parameter towards the faint end of the field luminoisty function. Our results also suggest that the properties of the Milky Way satellites are not shaped by starbursts, which are rare in this regime due to the relatively few mergers experienced by such galaxies in their histories \citep[e.g.,][]{Deason2022}.

\subsection{Higher-order correlations between parameters}
\label{sec:sobol_ST}

In the previous subsection, we have seen how the first-order Sobol' index, $S_1$, can be used to understand the primary physical processes that shape the number counts and chemical abundances of Milky Way satellite galaxies. Clearly, this cannot be the complete story: the final-day properties of the galaxy population will be the outcome of many physical processes conspiring {\it together}, as opposed to individual processes operating one-by-one. To this end, we will now consider the total sensitivity index, $S_\mathrm{T}$, defined in Equation~\ref{eq:ST_eq}, to understand the {\it joint} influence of these parameters on our model predictions.

\begin{figure*}
    \centering
    \includegraphics[width=0.49\linewidth]{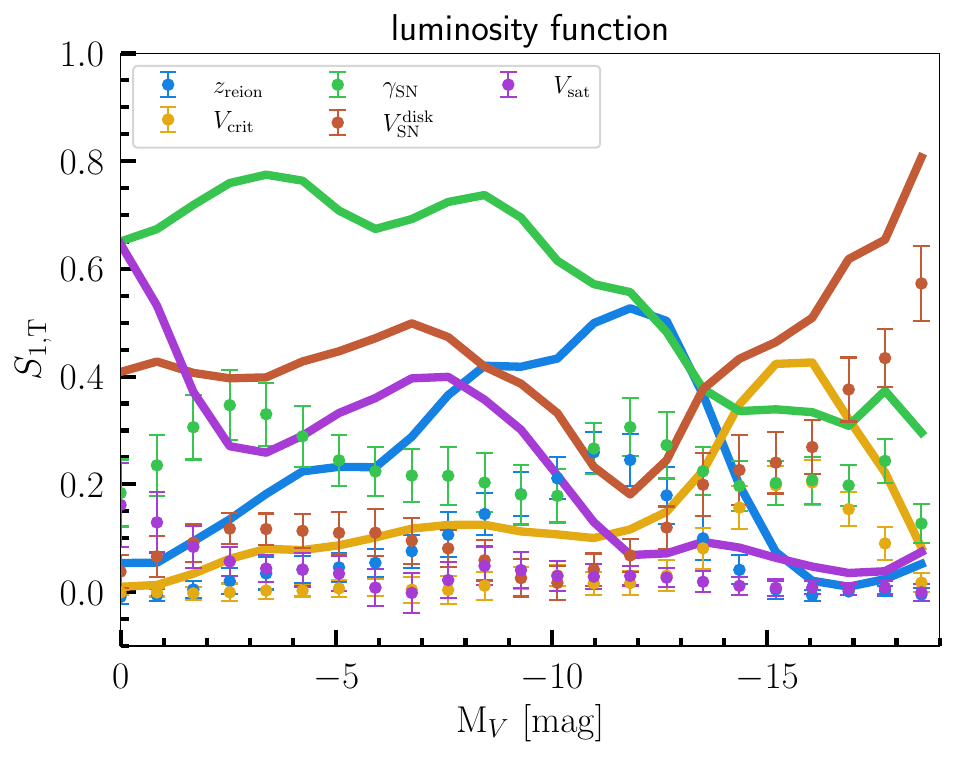}
    \includegraphics[width=0.49\linewidth]{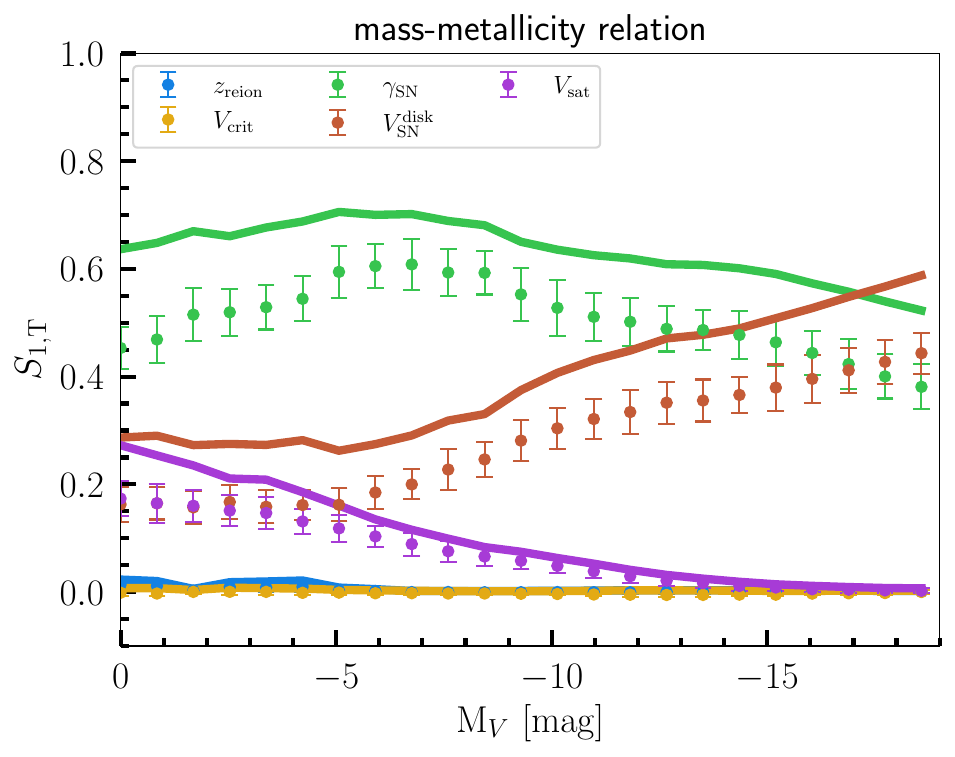}
    \caption{The first-order ($S_1$, shown as points with error bars) and total ($S_\mathrm{T}$, shown using solid lines) Sobol' sensitivity indices for various \galform{} input parameters. For clarity, we show results for only the five most influential parameters in the model (based on the tests presented in Figures~\ref{fig:S1_SLF} and~\ref{fig:S1_SMZ}). {\bf Left panel}: the sensitivity of the satellite luminosity function to these parameters. As noted previously, the parameters defining the supernova feedback scheme ($\gamma_\mathrm{SN}$ and $V_{\mathrm{SN}}^{\mathrm{disk}}$) are the most influential, although we observe considerable higher-order interaction with other parameters, as seen by the difference between the $S_1$ and $S_\mathrm{T}$ values of a given colour. {\bf Right panel}: as in the left-hand panel, but now for the mass-metallicity relation for Milky Way satellites. In contrast to the luminosity function, the predicted metallicities are almost entirely governed by the physics of supernova feedback, which shows comparatively little higher-order interactions with other physical processes in shaping these predictions.}
    \label{fig:ST_both}
\end{figure*}

This is shown in Figure~\ref{fig:ST_both}, where the total Sobol' indices for both the satellite luminosity function (left) and the mass-metallicity relation (right) are displayed. In each panel, the points represent the first-order index, $S_1$, and the associated error bars indicate the 95\% confidence interval. The solid line of each colour shows the total sensitivity index, $S_\mathrm{T}$, which encompasses the first and all higher-order interactions between each parameter and the rest\footnote{Note that for this reason, the sum total of the $S_\mathrm{T}$ values across all parameters in a given bin can be larger than 1.}. For clarity, we choose to show only the five most influential parameters of our model based on the analysis in Section~\ref{sec:sobol_S1}.

Consistent with what we have seen in Figures~\ref{fig:S1_SLF} and~\ref{fig:S1_SMZ}, the parameters defining the supernova feedback model ($\gamma_\mathrm{SN}$, $V_\mathrm{SN}^\mathrm{disk}$, and $V_\mathrm{sat}$) are the most dominant across all bins for both properties. Comparing the $S_1$ and $S_\mathrm{T}$ values, however, reveals some interesting new features. For instance, in the case of the luminosity function, there are large differences between the $S_1$ and $S_\mathrm{T}$ values for each of these parameters, suggesting that while they themselves have a sizeable first-order impact on the predicted number counts, they also exhibit significant correlations with other parameters in shaping the final prediction. Indeed, the increasing divergence of the $S_1$ and $S_\mathrm{T}$ values for $\mathrm{M}_V\gtrsim-13$ suggests that {\it the faint end of the luminosity function is the outcome of a more intricate set of interactions between physical processes in the model than the bright end}. 

Another interesting observation from Figure~\ref{fig:ST_both} is the behaviour of the reionisation parameters, $z_\mathrm{reion}$ and $V_\mathrm{crit}$ shown, respectively, using blue and gold. While their first-order sensitivity indices are only modest, their combined higher-order responses become significant. This is true especially for the redshift of reionisation, $z_\mathrm{reion}$. This may be understood as follows. Consider a model in which the value of $V_\mathrm{crit}$ is low ($\lesssim 20-30$ kms$^{-1}$), meaning that reionisation impacts only relatively small haloes ($M_\mathrm{peak}\lesssim10^9\,\mathrm{M}_\odot$; see Figure~\ref{fig:mstar-mhalo}). If the parameters defining the supernova feedback model (which, we recall, is the {\it dominant} physical process) are chosen such that the mass loading is not very large in this regime, galaxies are able to continue converting gas into stars, and shift to the brighter bins of the luminosity function, emptying out the fainter bins in the process. In this case, changing $z_\mathrm{reion}$ will have little to no effect (i.e., a low $S_1$), since its impact is in a regime (determined by the filtering scale, $V_\mathrm{crit}$) where there are few galaxies to begin with. This example illustrates how the physics of reionisation conspires with the physics of supernova feedback to shape the luminosity function, particularly at the faint end. For this reason, it is important to consider not just the impact of changing parameters individually, but in conjunction with others with which they exhibit correlations. This is the great advantage of using Sobol' indices for this kind of variance-based sensitivity analysis over one-at-a-time parameter variation techniques.

The right-hand panel of Figure~\ref{fig:ST_both} suggests that the picture is somewhat more straightforward in the case of the mass-metallicity relation. Here, it is only the parameters of the supernova feedback model that have any discernible effect on the model predictions. Furthermore, we also note that the difference between the $S_1$ and $S_\mathrm{T}$ values for each parameter is much smaller than it was in the case of the luminosity function, suggesting that {\it it is the first-order influence of the supernova feedback parameters that largely determines the variance in the predictions for galaxy metallicites} \citep[see also][]{Finlator2008,Pandya2021,Rey2025,VWheeler2025}.

\begin{figure*}
    \centering
    \includegraphics[width=0.49\linewidth]{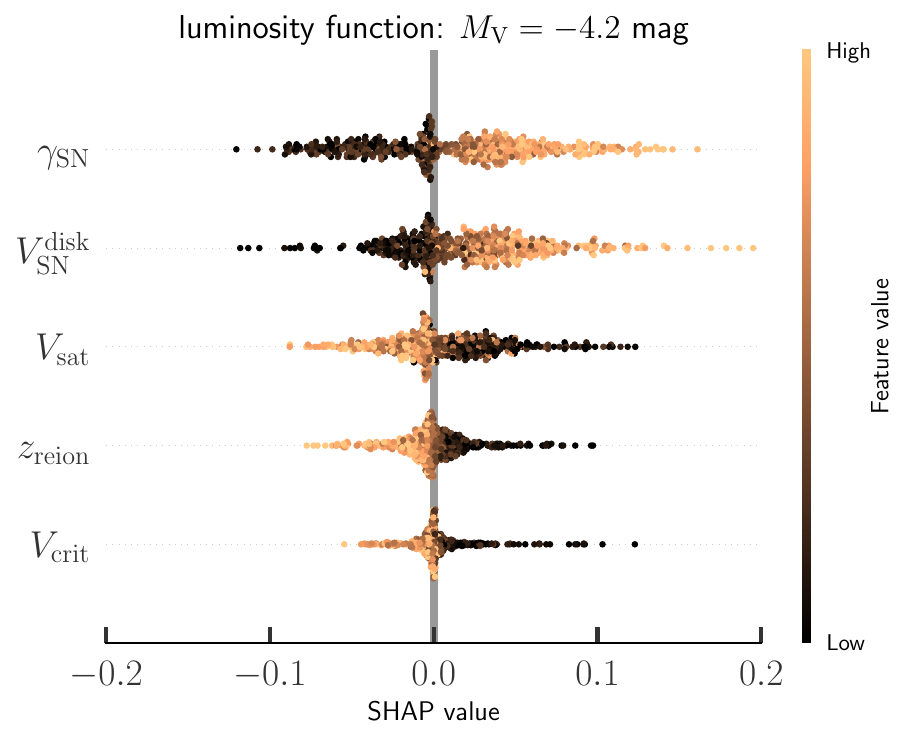}
    \includegraphics[width=0.49\linewidth]{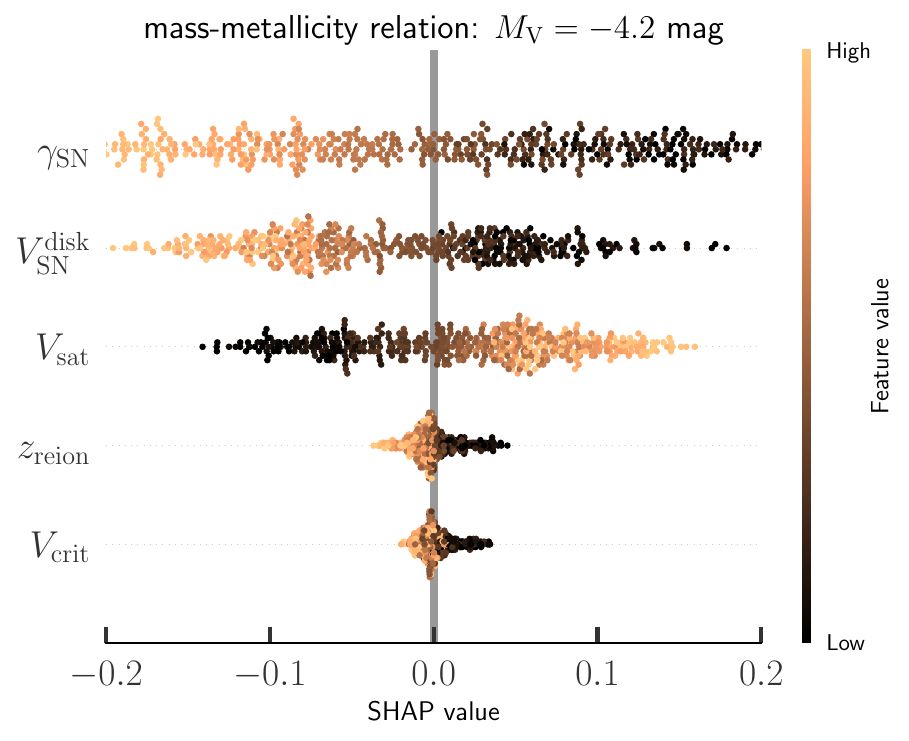} 
\\\includegraphics[width=0.49\linewidth]{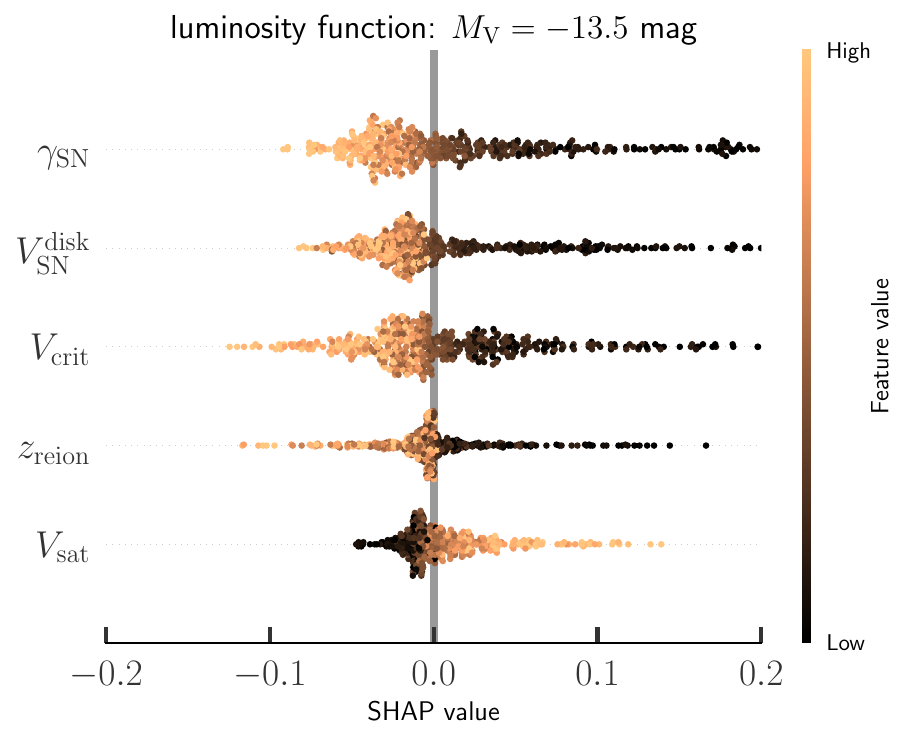}
\includegraphics[width=0.49\linewidth]{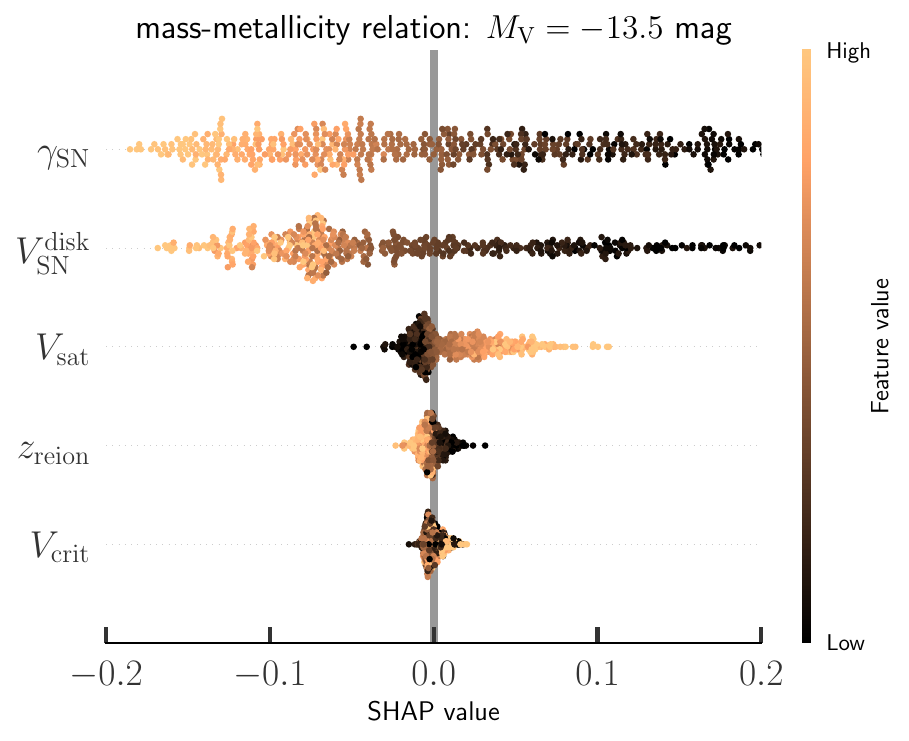}
\caption{`Beeswarm' diagrams showing SHapley Additive exPlanations (SHAP) values (see text in Section~\ref{sec:shap}), which illustrate the contribution of each input parameter to the \galform{} emulator predictions for the satellite luminosity function ({\bf left column}) and the mass-metallicity relation ({\bf right column}). For each parameter, the distribution of points shows its {\it impact} on the model output (SHAP value), with the colour of each point indicating the {\it value of the parameter} from low (dark) to high (light). The figure is split in order to separate the effect in the regime of faint satellites ({\bf upper row}, focussing on a bin at $\mathrm{M}_V=-4.2$ mag) from those of the brighter satellites ({\bf lower row}, focussing on a bin at $\mathrm{M}_V=-13.5$ mag). For clarity, we show only the five most influential parameters in our model, and they are rank-ordered in importance from top to bottom. The dominance of the supernova feedback parameters (the top three in the list) in shaping the mass-metallicity relation is again apparent, whereas reionisation plays a more significant role in determining number counts. Note also that varying the supernova feedback parameters from low to high pushes SHAP values for the luminosity function and mass-metallicity relation in {\it opposite directions} (i.e., the impact of these parameters is inverted for the two statistics) in the faint regime; this is not the case for brighter satellites.}
    \label{fig:SHAP_all}
\end{figure*}

The results of this section show that it is important not just to understand {\it which} parameters are influential, but also {\it how} they shape a particular statistic (i.e., the direction of the change). This will be the focus of the next subsection.

\subsection{The magnitude and directionality of parameter influences}
\label{sec:shap}

In order to quantify how (and by how much) each parameter influences the predictions of the \galform{} model, we use an approach from game theory, known as SHapley Additive exPlanations \citep[SHAP,][]{Lundberg2017}, which can be applied generally to explain the output of any machine learning model. In short, SHAP assigns each input feature (i.e., each physical parameter of the galaxy formation model) an importance value for a particular prediction. This `SHAP value' represents the contribution of that feature to perturbing the model prediction from a baseline value to its final predicted value. For a single prediction, $f(x)$, the SHAP explanation is given by:
\begin{equation} \label{eq:shap_eq} 
f(x) \approx g(x') = \phi_0 + \sum_{i=1}^{k} \phi_i x'_i,
\end{equation}
where $g$ is the explanation model, $x' \in \{0, 1\}^k$ is a simplified binary input representing whether a feature is present or absent, $k$ is the number of input features, and $\phi_i \in \mathbb{R}$ is the SHAP value for feature $i$. The term $\phi_0$ is the base value, which is the average model output over the entire training dataset (i.e., the prior) and can be estimated using the expectation value $E\left(f(x)\right)$. The SHAP value, $\phi_i$, for a feature $i$ is calculated by: 
\begin{equation} \label{eq:shap_eq_2}
\phi_i = \sum_{S \subseteq \mathbf{F} \setminus \{i\}} \frac{|S|!(|\mathbf{F}| - |S| - 1)!}{|\mathbf{F}|!} [f_x(S \cup \{i\}) - f_x(S)],
\end{equation}
where $\mathbf{F}$ is the set of all input features, $|\mathbf{F}|$ is the total number of features, and $|S|$ is the total number of features in the subset $S$. The term $f_x(S)$ is the expectation value of the model output conditional on the set of feature values in $S$. Equation~\ref{eq:shap_eq_2} computes the contribution of feature $i$ by considering every possible ordering of features and averaging their marginal contributions. The term in the square brackets represents the marginal contribution of feature $i$ to the prediction when it is added to the coalition of features $S$. To compute the SHAP values using our \galform{} emulator, we make use of the \texttt{SHAP}\footnote{\url{https://shap.readthedocs.io/}} library. The computation of SHAP values is somewhat expensive; in what follows, we evaluate Equation~\ref{eq:shap_eq_2} for a subset of 500 parameter combinations drawn from the Saltelli sampling.

\begin{figure*}
    \centering
    \includegraphics[width=\linewidth]{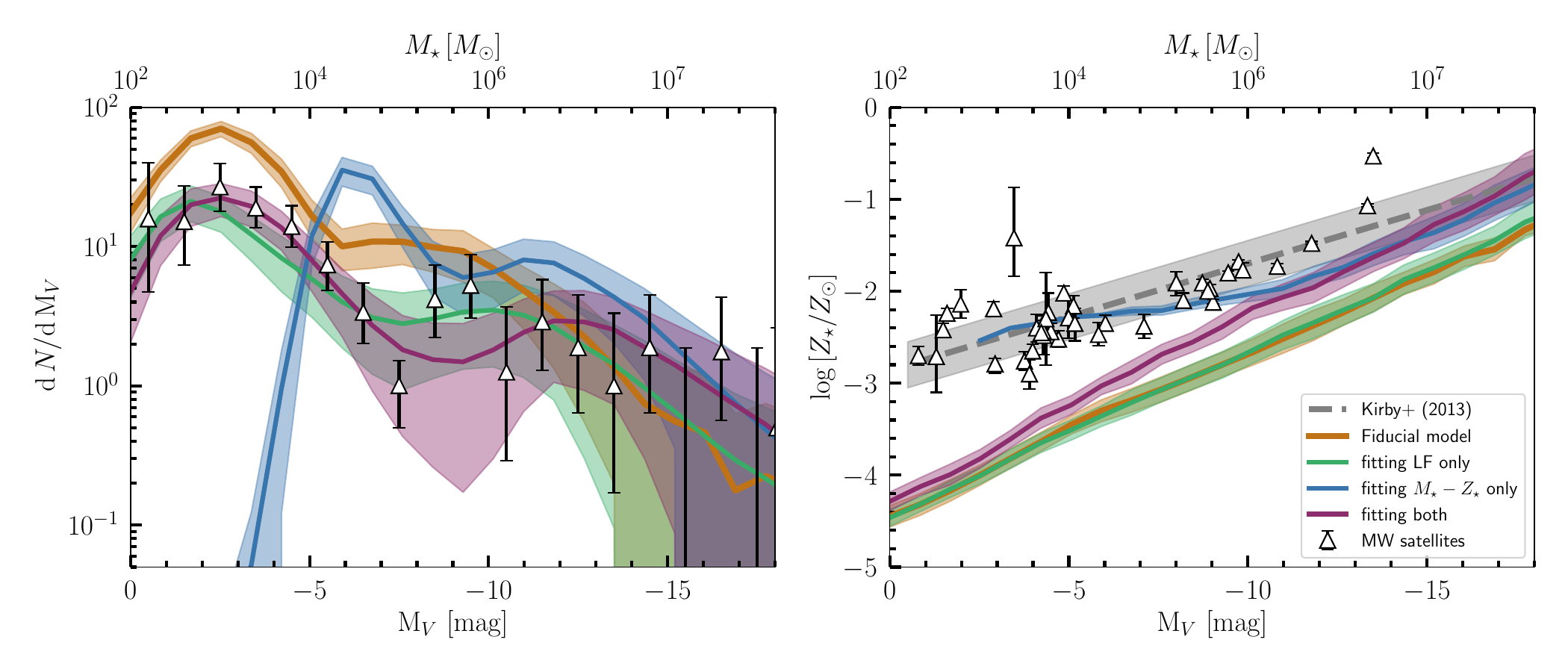}
    \caption{An illustration of the tension between simultaneously fitting the satellite luminosity function ({\bf left panel}) and the mass-metallicity relation ({\bf right panel}). The figure compares observational data (black triangles) against four distinct model parametrisations: the fiducial  model (copper curves), a model optimised solely to fit the luminosity function (green curves), a model optimised solely for the mass-metallicity relation (blue curves), and a model optimised to fit both data sets simultaneously (purple curves). We find that while the model optimised for the luminosity function closely reproduces the observed satellite counts, it predicts stellar metallicities that are substantially lower than those observed. Conversely, the model optimised for the mass-metallicity relation successfully matches the observational data in the right panel, but at the cost of a large discrepancy in the luminosity function, depleting the faint magnitude bins significantly and overpredicting the number of brighter satellites in the process. The model that aims to fit both data sets captures the shape of the luminosity function and does a better job in reproducing the metallicities of the brighter galaxies, but it still vastly underpredicts the metallicities of the faint satellites. This result indicates a fundamental conflict within the physical framework of the \galform{} model, as the parameters required to regulate satellite number counts are inconsistent with those needed to regulate their chemical abundances.}
    \label{fig:individual_fits}
\end{figure*}

Figure~\ref{fig:SHAP_all} shows the result of this analysis, where we focus our attention on two specific bins in each of the luminosity function (left column) and the mass-metallicity relation (right column): one in the regime of the ultrafaints, centred at $\mathrm{M}_V=-4.2$ mag (top row), and one in the regime of brighter satellites, centred at $\mathrm{M}_V=-13.5$ mag (bottom row). Each dot represents the result from a SHAP experiment, with the colour scale from black to copper indicating the adjustment of each parameter from low to high values. The location of each dot along the horizontal axis indicates whether this adjustment decreases (negative values) or increases (positive values) the output predicted by the model for that statistic. The density of points at a given SHAP value represents the frequency of that outcome\footnote{In these `beeswarm' diagrams, the values are arranged so that no two points overlap, which in turn gives the figure its distinctive `swarming' look.}. For simplicity, we limit our investigation to only the five most significant parameters. The parameters are listed from top to bottom in decreasing order of their contribution to the model.

The most significant insight we can draw from this figure is the opposing effect of supernova feedback on the faint satellite population for the two sets of predictions. In this magnitude bin, parameter values that decrease the number of satellites (negative SHAP values in the top-left panel) simultaneously drive up their metallicity (positive SHAP values in the top-right panel), revealing a fundamental tension in \galform{}'s ability to reproduce both statistics simultaneously. For the luminosity function, low values of $\gamma_\mathrm{SN}$, $V_\mathrm{SN}^\mathrm{disk}$ and $V_\mathrm{sat}$ push the SHAP values negative, thereby decreasing the predicted number of satellites. This might seem counter-intuitive as it suggests that reducing the strength of feedback also reduces the number of ultrafaints. However, this is to be understood not as a decrease in the absolute number formed but rather that in the absence of strong feedback, these galaxies are able to continue forming stars and become brighter, thereby shifting along the luminosity function and depleting the number counts in the fainter bins. In contrast, for the mass-metallicity relation, these same low values result in positive SHAP values: the metallicities of the ultrafaints increase when the supernova mass loading becomes weaker, as expected. 

This behaviour contrasts sharply with that of the brighter magnitude bin, shown in the bottom row. Here, the tension observed in the faint regime is no longer present. For example, high values of $\gamma_\mathrm{SN}$ now yield positive SHAP values for both the luminosity function and the mass-metallicity relation (which can now be understood following the natural conclusion of the arguments just presented for the fainter magnitude bin). This demonstrates that the impact of these physical parameters is not uniform across galaxy populations of different luminosities. The inability of our model to simultaneously match the luminosity function and the mass-metallicity relation (c.f. Figure~\ref{fig:all_models}) is therefore most acute for the faintest satellite galaxies.

\begin{table}
    \centering
    \renewcommand{\arraystretch}{1.5}
    \begin{tabularx}{\columnwidth}{ l  X  X X}
    \toprule
      {\bf Parameter} & {\bf Fitting LF only} & {\bf Fitting $M_\star$-$Z_\star$ only} & {\bf Fitting both}  \\
    \toprule
        $z_\mathrm{reion}$ & $7.34^{+0.85}_{-0.76}$ & $9.10^{+4.15}_{-4.59}$ & $8.52^{+0.48}_{-0.48}$ \\
        $V_\mathrm{crit}$ [kms$^{-1}$] & $42.16^{+2.71}_{-4.33}$ & $34.44^{+17.29}_{-18.52}$ & $41.85^{+2.25}_{-2.14}$\\
        $\gamma_\mathrm{SN}$ & $3.69^{+0.74}_{-0.41}$ & $2.51^{+1.13}_{-0.65}$ & $3.87^{+0.46}_{-0.46}$ \\
        $V_\mathrm{SN}^\mathrm{disk}$ [kms$^{-1}$] & $247.56^{+69.54}_{-73.23}$ & $296.97^{+157.23}_{-122.60}$ & $151.29^{+49.63}_{-28.45}$ \\
        $V_\mathrm{sat}$ [kms$^{-1}$] & $11.64^{+5.80}_{-4.43}$ & $32.58^{+23.83}_{-16.89}$ & $6.05^{+1.54}_{-0.79}$ \\
    \bottomrule
    \end{tabularx}
    \caption{The best-fit parameters for the MCMC exercise performed in Figure~\ref{fig:individual_fits}. The first column lists the parameters, while the second, third, and fourth columns, present the best-fit values (with $16^{\mathrm{th}}$ and $84^{\mathrm{th}}$ percentile confidence regions) when fitting to the luminosity function (LF), the mass-metallicity ($M_\star$-$Z_\star$) relation, and both statistics, respectively. It is clear that the luminosity function data prefer stronger supernova feedback, while the mass-metallicity relation pulls the parameter space into the regime of weaker supernova feedback.}
    \label{tab:best_fits}
\end{table}

A consequence of these behaviours is demonstrated in Figure~\ref{fig:individual_fits}, where we use our emulator to fit the observational data using MCMC \citep[for which we have used the \texttt{emcee}\footnote{\url{https://emcee.readthedocs.io}} package,][]{emcee2013}. In this figure, the green curve shows the result of fitting to the observed satellite luminosity function only, whereas the blue curve shows the result of fitting to the observed mass-metallicity relation only. The copper curves in each panel show the predictions of the fiducial model, which has been included for reference. 

The tension outlined earlier in this section is now seen quite clearly: the model that is optimised to reproduce the satellite number counts vastly underpredicts the galaxy metallicities. Conversely, the model optimised to fit the mass-metallicity relation completely fails to reproduce the shape of the luminosity function, particularly in the regime of the ultrafaints. Our investigations in Figures~\ref{fig:S1_SLF}-\ref{fig:ST_both} help us understand why: the relatively strong feedback needed to prevent faint galaxies from becoming too massive results in metals being blown out of galaxies too strongly. On the other hand, reducing the strength of feedback to remedy this has the effect of reducing the number counts of faint satellites at present-day, and overpredicting the number of bright satellites in turn. The best-fit parameters obtained from each of these tests are listed in Table~\ref{tab:best_fits}. Indeed, the case where the model is optimised to fit the observed number counts prefers much weaker supernova feedback than the case where the model is fit to the mass-metallicity relation. In the latter case, the parameter $V_\mathrm{sat}$ is pushed higher; this is driven by the attempt to fit the metallicites of the faintest satellites, which in turn results in a break in the slope of this relation at $\mathrm{M}_V\approx-10$ mag ($M_\star\approx10^6\,\mathrm{M}_\odot$). This model also predicts essentially no satellites with $\mathrm{M}_V\gtrsim-3$ mag ($M_\star\lesssim10^3\,\mathrm{M}_\odot$), as the galaxies that would have populated these magnitude bins are allowed to become brighter due to the reduced efficiency of supernova feedback. We also note that this effect also shifts the `kink' in the luminosity function to brighter magnitudes than in the fiducial model, as the effective stellar mass (luminosity) of haloes that are affected by reionisation is now larger. We also note that in this case, some of the parameters are rather poorly constrained; this also suggests that the satellite luminosity function is a more informative statistic for constraining the parameters of a galaxy formation model than the mass-metallicity relation (each taken individually).

An obvious next question is to ask if there is {\it any} combination of parameters that allows us to fit both sets of data simultaneously. In fact, we find that the answer to this is no: a joint MCMC fitting exercise does not identify a particular combination of parameters (across the entire 11-dimensional space) that provides a good fit to both sets of data. The model that best fits both sets of data is shown in Figure~\ref{fig:individual_fits} in purple, with the best-fit parameters listed in the final column of Table~\ref{tab:best_fits}. We find that the luminosity function is well reproduced in both its shape and normalisation, and the metallicities of the brighter galaxies are now in better agreement with the data. However, there is still a large discrepancy at the faint end, where the feedback is still too strong in order to keep the number counts in agreement. This reveals a fundamental limitation of the present model, and suggests that it is lacking a key physical process (if not more). In the next section, we consider some possible examples.

\section{Discussion}
\label{sec:discussion}

In this section, we consider possible extensions to the standard \galform{} model that could be used to address the inability of the current model to reproduce the (observed) number counts and stellar metallicites of Milky Way satellites simultaneously.

\subsection{Supernova yields and the IMF}
\label{sec:yields_imf}
The most obvious way to enhance the metallicities of the satellites in our model is to increase the yield, $p$, which enters the system of equations governing chemical evolution in our model in Equation~\ref{eq:yield}. Indeed, this parameter is not without uncertainty, and its value depends on a variety of aspects related to stellar evolutionary models, such as knowledge of nuclear reaction rates and uncertainties in modelling supernova explosion mechanisms \citep[e.g.,][]{Romano2010,Nomoto2013,Vincenzo2016}. This number cannot be changed arbitrarily, however: the yield per stellar generation (in addition to the recycling fraction, $R$) can be estimated self-consistently given a choice of IMF. One option may be to opt for a different IMF -- switching from the \cite{Kennicutt1983} IMF (which is the default assumption in \galform{}) to the one from \cite{Chabrier2003} enhances the yields by roughly a factor of two. However, this is substantially smaller than the factor of $\approx30$ discrepancy we see in the model (Figure~\ref{fig:all_models}), so this is unlikely to be the sole explanation.

\begin{figure*}
    \centering
    \includegraphics[width=\linewidth]{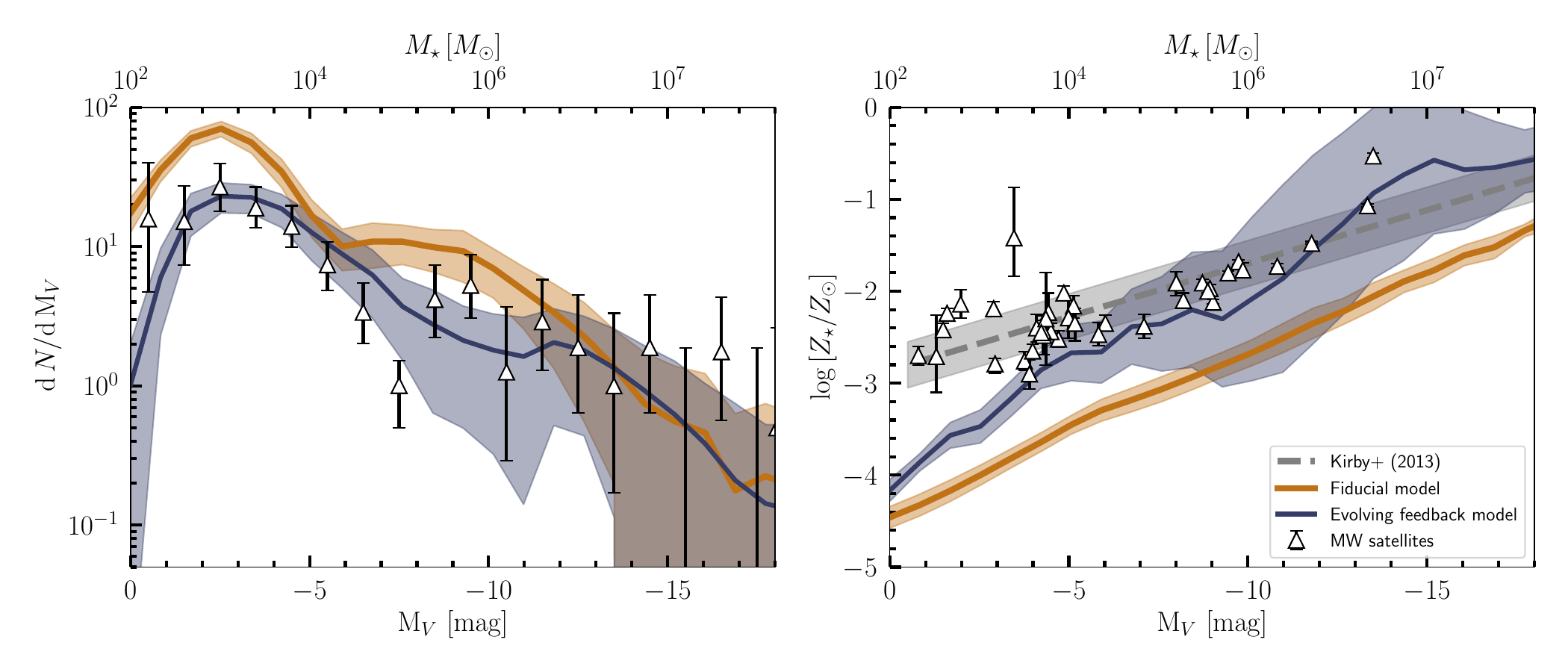}
    \caption{As Figure~\ref{fig:individual_fits}, with the navy blue curves representing the predictions of the best-fitting model in which the strength of supernova feedback is allowed to evolve with redshift (see Section~\ref{sec:evolve} for details).}
    \label{fig:evolving_fits}
\end{figure*}

\subsection{Metal loading in winds}
\label{sec:Z_loading}
An appealing alternative involves assuming that mass and metals are ejected out of galaxies into the circumgalactic medium (and beyond) with different efficiencies. This selective enrichment of galactic outflows (relative to the ISM) is parametrised in the form of a separate `metal loading' factor, $\zeta$, defined as the ratio between the metallicity of the wind and that of the ISM. The introduction of such a parameter modifies Equation~\ref{eq:yield} subtly:
\begin{equation}
    \dot{M}_\mathrm{cold}^Z = Z_\mathrm{hot}\dot{M}_\mathrm{acc} + \left[p - \left(1-R+{\color{teal}{\mathbf{\zeta}}}\beta\right)Z_\mathrm{cold}\right]\psi.
\end{equation}
The model presented here effectively assumes ${\color{teal}{\mathbf{\zeta}}}=1$, but there is evidence from galaxies observed locally that this number could vary \citep[e.g.,][]{Peeples2011,Chisholm2018}\footnote{In terms of the notation used in this paper, \cite{Peeples2011} define their wind loading as $\eta_w \equiv \zeta \beta$.}. Recent high-resolution simulations of galactic outflows ascribe the existence of a separate wind loading factor to the exchange and transport of metals in a multiphase outflow \citep[e.g.,][]{Vijayan2024}. In principle, the metal loading factor could vary as a function of galaxy properties or, indeed, the type of metals being loaded. While these details are beyond the scope of the present investigation, it is clear that the inclusion of such a term could provide our model with an extra degree of freedom that can be used to fit the model to the two sets of properties somewhat more independently.

\subsection{Delayed enrichment from massive stars}
\label{sec:delayed}
The chemical evolution network in \galform{} (described in Section~\ref{sec:chemistry}) assumes the instantaneous recycling approximation throughout, in which a fixed metal yield is ejected immediately after a star formation event. Under this approximation, all stars with masses $\gtrsim 1\,\mathrm{M}_\odot$ are considered to die instantaneously, while less massive stars are assumed to have infinite lifetimes. This approximation is reasonable in the case of elements produced by short-lived stars (such as oxygen). In observations, on the other hand, the chemical abundances of Milky Way satellites are predominantly based on iron, for which the production source is long-lived, massive stars. In this scenario, the instantaneous recycling approximation is no longer valid. Using the \textsc{l-galaxies} semi-analytic model, \cite{Yates2013} developed a detailed chemical evolution network that includes delayed enrichment from stellar winds, Type Ia and Type II supernovae, as well as metallicity-dependent yields. While this gradual enrichment provides a better match to observed stellar metallicities in the regime they consider ($M_\star\gtrsim10^9\,\mathrm{M}_\odot$), the actual shift in the normalisation of this relation is modest (around 0.1-0.2 dex). Accounting for delayed enrichment, therefore, is unlikely to be the fix on its own. 

\subsection{Preventive feedback}
\label{sec:preventive}
\citet{Lu2017} demonstrated that a combination of `preventive' feedback (which prevents gas from accreting onto low-mass haloes in the first place) and `ejective' feedback can simultaneously match the luminosity function and the mass-metallicity relation. Furthermore, this mechanism is able to reproduce the observed flatness of the [Fe/H] distribution at the low-mass end of the satellite population. Implementing a preventive feedback scheme in \galform{}, perhaps linked to the thermodynamic state of the halo gas or pre-heating from reionisation, could offer an alternative solution to the tension we identify. We leave the examination of an additional feedback mode of this kind to future work.

\subsection{Redshift-dependent supernova feedback}
\label{sec:evolve}
A previous investigation into the joint predictions for the abundance and metallicities of Milky Way satellites was undertaken by \cite{Hou2016}, albeit without consideration of the ultrafaint regime. These authors introduced a new model of supernova feedback in \galform{}, in which the feedback strength is allowed to vary as a function of redshift. In particular, this model behaves similarly to the fiducial \cite{Lacey2016} model at low redshift ($z\leq4$), but is weaker at earlier times. Specifically, the parameter $V_\mathrm{SN}^\mathrm{disk}$ evolves with redshift, and the power-law index of the mass loading factor also becomes shallower below some threshold circular velocity. 

Although such a model might initially appear somewhat contrived, it may not be unreasonable to expect the efficiency of supernova feedback to evolve over time. In our fiducial setup, the mass loading factor depends only on the gravitational potential well of the galaxy (parametrised by the circular velocity, $V_c$; see Equation~\ref{eq:loading}) when, in reality, this may also depend on the gas density, gas metallicity, and molecular gas fraction in the galaxy \citep[see, e.g.,][]{Lagos2013}. This is because the gas density and metallicity in the ISM determine the local gas cooling rate. This rate, in turn, dictates the fraction of the energy injected by supernovae that can eventually be used to launch outflows. However, dense molecular gas in galaxies may not be directly affected by supernova explosions and therefore may not be ejected as outflows. This complex dependence of the mass loading factor on the surrounding gas conditions may not be captured in terms of a simple scaling in $V_c$ only, especially when considered across a wide range of redshifts.

Figure~\ref{fig:evolving_fits} shows the result of fitting the evolving feedback model to the satellite luminosity function and mass-metallicity relation simultaneously. In performing this fit, we have allowed the free parameters of the model to vary from the best-fit values listed in \cite{Hou2016}; their exact values are not important as the aim of the present exercise is not to find the `right' model, but to demonstrate that one can indeed be identified. This model, with its extra freedom to evolve feedback strength over time, does a much better job than the fiducial model (shown in copper) showing, in particular, a far better match to the observed mass-metallicity relation. The evolving feedback model still underpredicts the metallicities of galaxies with $\mathrm{M}_V\gtrsim-5$ mag. It is interesting to note that this model also predicts a much larger scatter in the metallicities of bright satellites than our fiducial model. 

\begin{figure}
    \centering
    \includegraphics[width=\columnwidth]{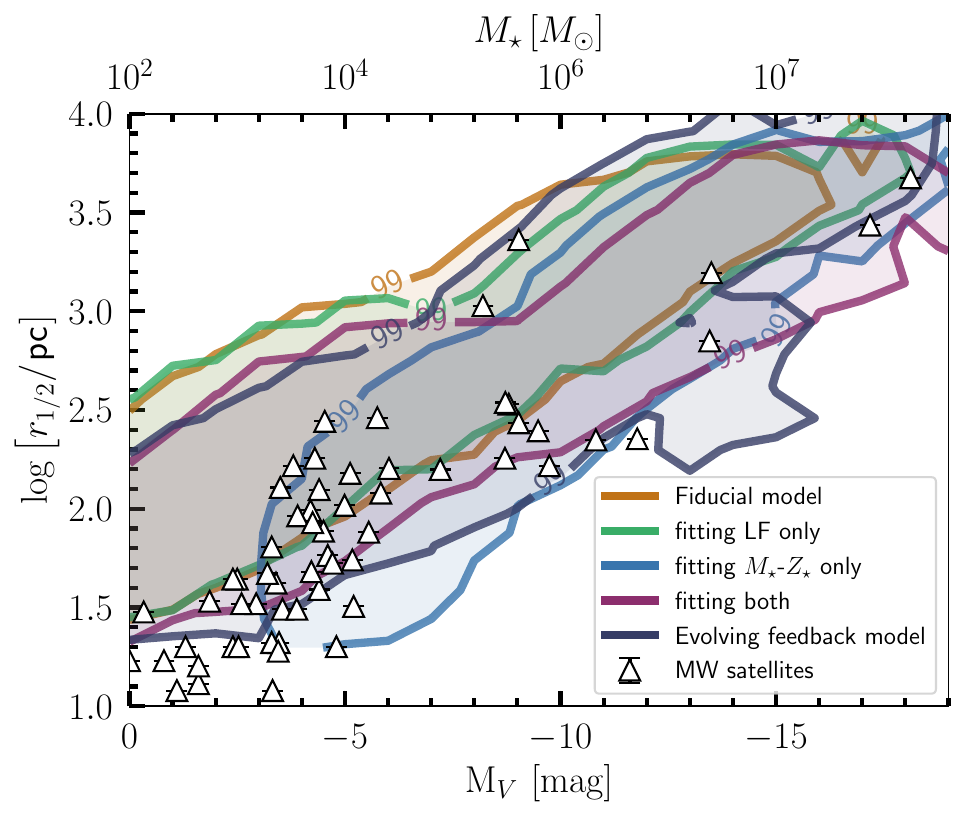}
    \caption{A comparison of the mass-size relation for Milky Way satellites. We compare the predictions of five different \galform{} variations: the fiducial model (copper), models optimised to fit only the luminosity function (green), only the mass-metallicity relation (blue), or both (purple), and a model with a redshift-dependent feedback efficiency (navy blue). In each case, we show contours that encompass 99\% of the population predicted by each model. Observational data compiled by \citet{Drlica2020} are shown using black triangles.  While there is significant scatter, a clear trend emerges: models that are most discrepant in the mass-metallicity relation predict sizes that are most discrepant with the observational data (copper and green), whereas those in which the predicted metallicities are in better agreement with the data (blue and navy blue) are also in better agreement in terms of galaxy sizes.}
    \label{fig:compare_sizes}
\end{figure}

An interesting by-product of this exercise is that the best-fit evolving feedback model also predicts galaxy sizes that are in good agreement with the half-light radii measured from the observed Milky Way satellites (which, on average, are otherwise too large by a factor of $\approx3-5$ in the fiducial model). In fact, a general observation we make in our experiments is that models in which the feedback strength is adjusted to bring the predicted metallicities into closer agreement with the observed values tend to predict sizes that are also reasonably consistent with the data. This is shown in Figure~\ref{fig:compare_sizes}, where the models introduced in Figure~\ref{fig:individual_fits}, as well as the evolving feedback model, are used to predict the mass-size relation for Milky Way satellites. All models predict significant scatter in this relationship; the size of the scatter is not uniform across models. The model that is fit to the luminosity function only (green) predicts systematically larger sizes than what is measured, whereas the one that is fit to the mass-metallicity relation only (blue) is generally in good agreement with the data. The other models lie somewhere in between these limiting cases.  None of the models are able to reproduce the very compact sizes of the faintest satellites, but note that we have not taken into account the reduction of satellite galaxy sizes due to tidal effects -- this could have a significant impact in this regime \citep[e.g.][]{Errani2022}. This result suggests that the underlying physics of feedback that regulates the chemical abundances of galaxies may play an equally influential role in regulating their characteristic sizes.

\section{Conclusions}
\label{sec:conclusions}
We have performed a systematic investigation to understand which physical processes are the most influential in shaping the properties of the Milky Way satellite galaxy population. To this end, we use a semi-analytic model of galaxy formation, \galform{}, whose predictions are then used to train a neural network emulator to efficiently explore the 11-dimensional parameter space that defines the model (Table~\ref{tab:params}). The parameters of the \galform{} model are calibrated to reproduce a small set of observations of the field galaxy population, on scales much larger than those of interest in this paper; forecasts made by the model in the dwarf regime can therefore be considered genuine predictions. Our focus in this paper is on two distinct statistics: the luminosity function of satellites, and their mass-metallicity relation (both measured at $z=0$). We employ a variance-based sensitivity analysis using Sobol' indices to determine the first- and higher-order influence of the model parameters in shaping these predictions. We then use SHapley Additive exPlanation (SHAP) values, a game theoretic approach, to quantify {\it how} each parameter impacts each of these model predictions. Our main findings are summarised below:
\begin{enumerate}
\item The abundance of bright satellites ($\mathrm{M}_V \lesssim -13$) is regulated almost exclusively by the physics of supernova feedback. Specifically, the power-law index of the mass loading factor ($\gamma_\mathrm{SN}$) and its normalisation in galactic disks ($V_\mathrm{SN}^{\mathrm{disk}}$) are the two most impactful parameters at first-order (Figure~\ref{fig:S1_SLF}).
\item The faint end of the satellite luminosity function is the result of a more complex interplay between physical processes. While supernova feedback remains crucial, the redshift of reionisation ($z_\mathrm{reion}$) and the filtering scale below which reionisation stalls gas cooling in haloes ($V_\mathrm{crit}$) become equally important. 
\item In contrast, the mass-metallicity relation of satellites is governed almost entirely by supernova feedback across all magnitude bins. The parameters defining the mass loading factor ($\gamma_\mathrm{SN}$, $V_\mathrm{SN}^\mathrm{disk}$, and $V_\mathrm{sat}$) are the only ones with any discernible influence. Reionisation has little or no impact on the metallicities of satellites at any mass scale (Figure~\ref{fig:S1_SMZ}).
\item The significant higher-order interactions between parameters indicate that the luminosity function at present day, particularly in the regime of faint satellites, is shaped by the combined effect of reionisation and supernova feedback.
On the other hand, the supernova mass loading parameters exhibit relatively few higher-order interactions when conspiring to shape the mass-metallicity relation (Figure~\ref{fig:ST_both}).
\item Using SHAP values, a fundamental tension within the \galform{} framework emerges: the parameter values required to reproduce the observed satellite luminosity function are inconsistent with those needed to match the observed mass-metallicity relation. Strong feedback, which is necessary to suppress star formation and prevent the overproduction of bright satellites, expels too many metals, resulting in stellar metallicities that are far too low. Conversely, weakening the feedback to retain metals results in a luminosity function with the incorrect shape (Figure~\ref{fig:SHAP_all}).
\item A consequence of this tension is that the model that best fits the observed luminosity function underpredicts stellar metallicities by more than an order of magnitude, while the model that best fits the mass-metallicity relation fails to reproduce the shape of the luminosity function. We are unable to find a combination of parameters within the fiducial framework that fits both data sets simultaneously (Figure~\ref{fig:individual_fits}).
\item We consider potential solutions to this tension, which includes options like updated stellar yields, delayed enrichment from long-lived stars, and separate metal loading in supernova outflows. The `true' solution might involve some combination of these possibilities. We also show how a model in which the mass loading factor is allowed to evolve in redshift is capable of fitting these data sets simultaneously (Figure~\ref{fig:evolving_fits}).
\item Finally, we find that models that show closer agreement with the observed mass-metallicity relation tend to also predict galaxy sizes that are in better agreement with the observed sizes of Milky Way satellites, essentially `for free' (Figure~\ref{fig:compare_sizes}). This suggests that the physical processes that determine one property are equally influential in shaping the other.
\end{enumerate}
A criticism that is sometimes levelled at semi-analytic models is that the large number of parameters used to define these models can be varied at will to fit any data set as desired. A further criticism is that this proliferation of parameters makes it hard to determine how a model makes any given prediction -- and, indeed, if the underlying reason(s) can be tied to physically-informed processes in the model. Our investigation shows that the use of techniques like variance-based sensitivity analysis and SHAP values can help turn an otherwise complicated, `black box'-like model into a more explainable framework. These methods allow us to visualise familiar statistics like luminosity functions in terms of parameter contributions to individual bins in the prediction, and can be used to quantify higher-order correlations between parameters. Perhaps more importantly, identifying sets of predictions that pull the parameter space in opposite directions allows us to pinpoint physical limitations within the model; this, in turn, opens up an opportunity to develop a more comprehensive, informed framework for modelling the formation and evolution of galaxies.

\section*{Acknowledgements}

We thank the anonymous referee for several useful recommendations that have improved the clarity of this paper. We are grateful to Carlton Baugh, John Helly, and Cedric Lacey for informative discussions during the course of this work. SB is supported by the UK Research and Innovation (UKRI) Future Leaders Fellowship [grant number MR/V023381/1]. AD is supported by a Royal Society University Research Fellowship. SB and AD acknowledge support from the  Science and Technology Facilities Council (STFC) [grant number ST/X001075/1]. This work used the DiRAC@Durham facility managed by the Institute for Computational Cosmology on behalf of the STFC DiRAC HPC Facility (\url{www.dirac.ac.uk}). The equipment was funded by BEIS capital funding via STFC capital grants ST/K00042X/1, ST/P002293/1, ST/R002371/1 and ST/S002502/1, Durham University and STFC operations grant ST/R000832/1. DiRAC is part of the National e-Infrastructure.

\section*{Data Availability}

The processed (binned) luminosity functions and mass-metallicity relations for the 2600 \galform{} runs that are used as the basis of this work can be accessed online at: \url{https://github.com/sownakbose/slf_emulator_project.git}.



\bibliographystyle{mnras}
\bibliography{refs} 




\appendix

\section{The host haloes of satellite galaxies}
\label{sec:gal_host}

\begin{figure}
    \centering
    \includegraphics[width=\columnwidth]{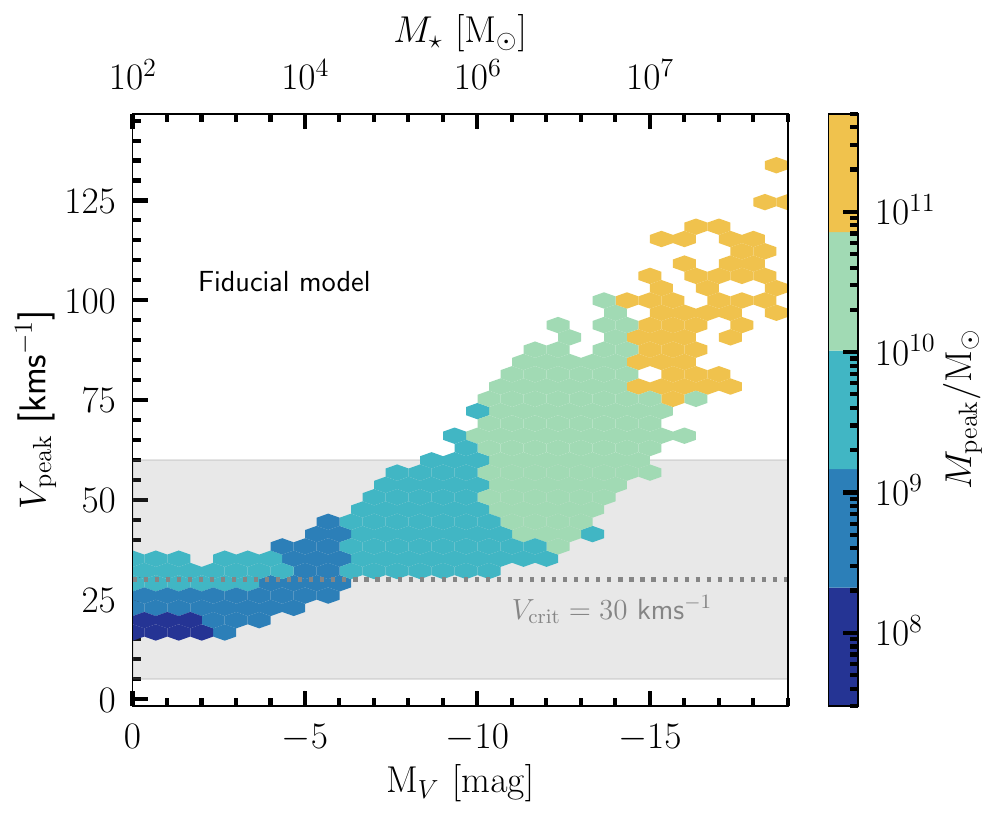}
    \caption{The distribution of galaxies in the fiducial \galform{} model in the plane of $V_\mathrm{peak}$ and $\mathrm{M}_V$. Here, $V_\mathrm{peak}$ is the peak circular velocity of the satellite measured along its merger tree. Each hexagonal bin is coloured according to the corresponding peak halo mass, as indicated by the colour bar. As expected, brighter, more massive galaxies (lower $\mathrm{M}_V$, higher $M_\star$) are associated with higher peak circular velocities and larger host dark matter haloes. There is a break in the slope of this relation, which is linked to the mass scale below which reionisation effects become important; this is denoted by the dotted horizontal line at $30$ kms$^{-1}$, corresponding to the choice of $V_\mathrm{crit}$ in the fiducial model. The grey shaded region marks the range over which this parameter is varied in Section~\ref{sec:main}.}
    \label{fig:mstar-mhalo}
\end{figure}

Figure~\ref{fig:mstar-mhalo} shows a variation of the stellar-to-halo mass relation for satellite galaxies in \galform{}. Here, we show only the results of the fiducial model as a reference. The lower and upper axes, respectively, show the (present-day) absolute V-band magnitude and stellar mass of the satellite, while the vertical axis shows its peak circular velocity, $V_\mathrm{peak}$, which has been measured along the main progenitor branch of the host subhalo. $V_\mathrm{peak}$, by definition, is reached before the galaxy becomes a satellite, typically in its final phases as an isolated, central galaxy. The colour of each hexagonal pixel indicates the corresponding peak halo mass, $M_\mathrm{peak}$. 

The relation shows a somewhat linear scaling at the bright end, with brighter, more massive satellites residing in more massive, higher $V_\mathrm{peak}$ haloes. There is a break in this relation at around $M_\mathrm{peak}=2\times10^{9}\,\mathrm{M}_\odot$, or $V_\mathrm{peak}=30$ kms$^{-1}$. This corresponds exactly to our choice for the filtering scale for reionisation in the fiducial \galform{} model, which is indicated using the dotted horizontal line. The net effect is a change in the slope of this relation, with the fainter galaxies ($\mathrm{M}_V\gtrsim-5$, $M_\star\lesssim10^4\,\mathrm{M}_\odot$) now residing in a much smaller range of host haloes (this would manifest as a {\it steepening} of the relation in the plane of $M_\star$-$M_\mathrm{peak}$). The bulk of the satellite population inhabits haloes in the range $16\lesssim V_\mathrm{peak}/\mathrm{kms}^{-1}\lesssim80$, where the lower limit is set by the fact that \galform{} galaxies only form inside haloes that lie above the hydrogen atomic cooling limit.

\section{Saltelli sampling}
\label{sec:saltelli}

\begin{figure*}
    \centering
    \includegraphics[width=\textwidth]{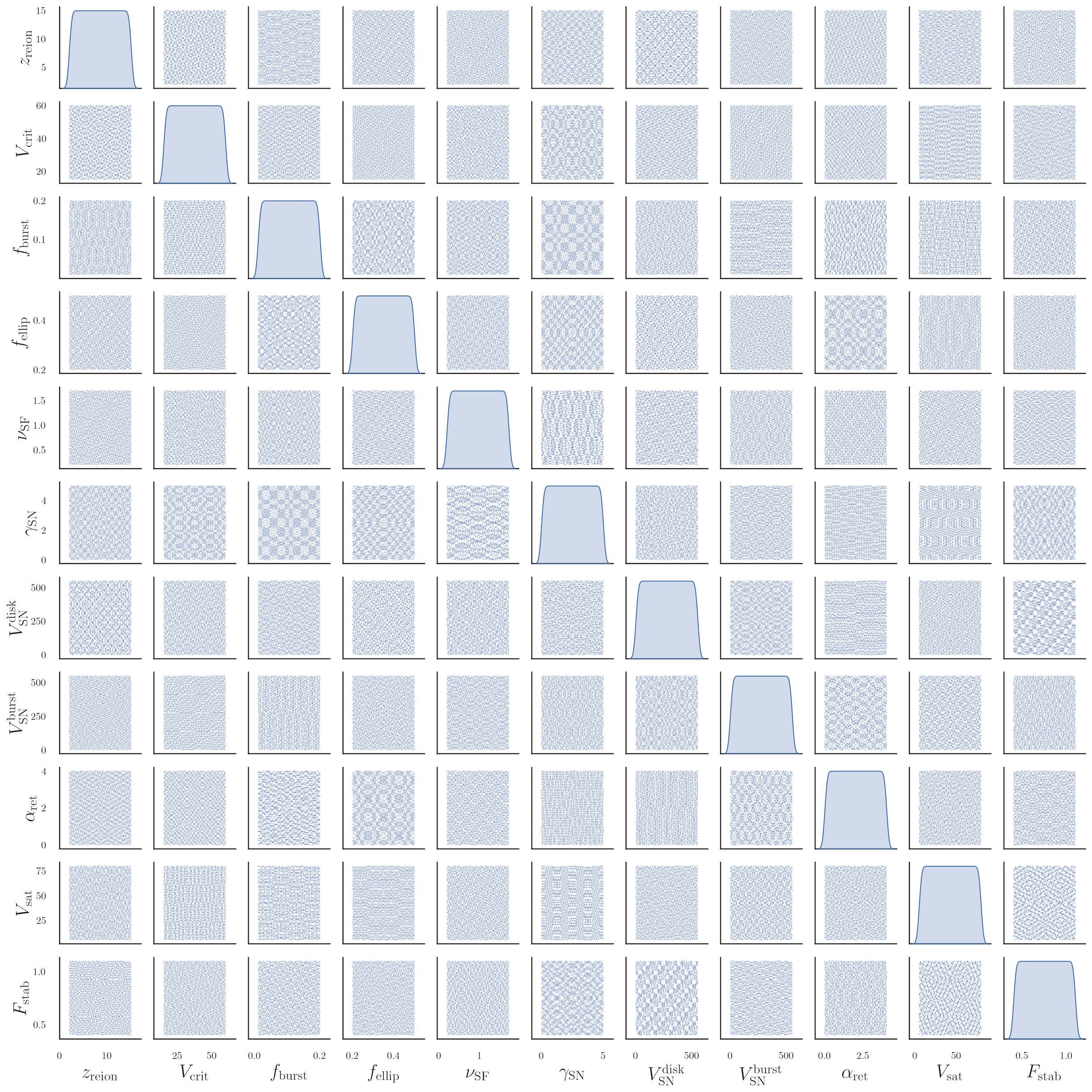}
    \caption{The distribution of the 11 input parameters for the \galform{} model for quantifying Sobol' sensitivity indices. The parameter set is generated using a Saltelli sampling sequence. The panels along the diagonal show the one-dimensional marginal probability distribution for each parameter, while the off-diagonal panels show the two-dimensional joint distributions for each pair of parameters. The structure noticeable in some panels are the consequence of projecting the parameter space into two dimensions. The marginal distributions along the diagonal are consistent with uniform sampling across the prior range, while the joint distributions in the off-diagonal panels show no apparent correlations between any pair of parameters; this suggests that the parameter space has been sampled in an unbiased manner.}
    \label{fig:saltelli_fig}
\end{figure*}

Computing Sobol' sensitivity indices (Section~\ref{sec:sobol}) efficiently requires a sampling strategy that is uniform, unbiased, and optimised across the high-dimensional input space. To this end, we make use of the Saltelli sampler \citep{Saltelli2002}, which is especially suited for our purpose. A brief description of this method is as follows:
\begin{enumerate}
    \item We begin by generating two independent sample matrices, $\mathbf{A}$ and $\mathbf{B}$, each of size $N \times k$, where $N$ is the number of samples and $k$ is the number of parameters. Each column corresponds to a parameter, and each row is a sample point in the $k$-dimensional parameter space.
    \item For each parameter $X_i$ (from $i=1$ to $k$), a new matrix, $\mathbf{A_B^i}$, is constructed. This matrix is identical to $\mathbf{A}$ except that its $i$-th column is replaced with the $i$-th column from matrix $\mathbf{B}$.
    \item Next, the model $f(\mathbf{X})$ is evaluated for each row in all the generated matrices: $\mathbf{A}$, $\mathbf{B}$, and the $k$ hybrid matrices $\mathbf{A_B^i}$. This results in a total of $N \times (k+2)$ model evaluations. In practice, this corresponds to evaluating our \galform{} emulators at $N \times (k+2)$ points in the input parameter space.
    \item Sobol' indices are estimated using the outputs from these evaluations. For example, the first-order index, $S_i$, is estimated as:
    \begin{equation}
        S_i \approx \frac{\frac{1}{N} \sum_{j=1}^{N} f(\mathbf{B})_j (f(\mathbf{A_B^i})_j - f(\mathbf{A})_j)}{\mathrm{Var}(Y)}
    \end{equation}
\end{enumerate}
Figure~\ref{fig:saltelli_fig} shows the distributions across our 11-dimensional input parameter space. The histograms along the diagonal (one-dimensional marginal distributions for {\it individual} parameters) and off-diagonal (two-dimensional joint distributions for {\it pairs} of parameters) panels showcase a uniform and unbiased sampling method. 


\bsp	
\label{lastpage}
\end{document}